\documentclass[sigconf]{acmart}
 
\usepackage{booktabs} %
\usepackage{tikz,subfig}
\usepackage{color,enumitem}
\usepackage{listings}             %
\usepackage{amssymb}
\usepackage{balance}
\usepackage{xcolor}
\usepackage{multirow}
\usepackage{tabularx}
\usepackage{array}
\newcolumntype{Y}{>{\centering\arraybackslash}X}

\newcommand{\hpara}[1]{\textcolor{blue}{[#1]}}
\renewcommand{\hpara}[1]{}

\usepackage{xspace}
\newcommand{\spin}{sPIN\xspace}
\setcopyright{acmcopyright}

\usepackage{tikz}
\usetikzlibrary{positioning,fit,calc,shapes.geometric}
\usepackage{soul} 

\definecolor{lightyellow}{RGB}{250, 250, 180}
\definecolor{HLYELLOW}{RGB}{240, 127, 0}
\definecolor{hlyellow}{RGB}{240, 127, 0}
\sethlcolor{lightyellow}

\usepackage[framemethod=tikz]{mdframed}

\makeatletter

\patchcmd\mdf@put@frame@i{\hrule \@height\z@ \@width\hsize\vfill}{}{}{}
\patchcmd\mdf@put@frame@i{\hrule \@height\z@ \@width\hsize\vfill}{}{}{}
\patchcmd\mdf@put@frame@i{\hrule \@height\z@ \@width\hsize\vfill}{}{}{}

\makeatother

\mdfsetup{backgroundcolor=grisframe,
    hidealllines=true,
    backgroundcolor=lightyellow,
    innerleftmargin=3pt,
    innerrightmargin=3pt,
    innertopmargin=2pt,
    innerbottommargin=1pt,
    leftmargin=-3pt,
    rightmargin=-3pt,
    splitbottomskip=0
}

\author{Salvatore Di Girolamo$^{1,4}$, Konstantin Taranov$^1$, Andreas Kurth$^2$, Michael Schaffner$^2$, Timo Schneider$^1$, Jakub Ber\'anek$^3$, Maciej Besta$^1$, Luca Benini$^2$, Duncan Roweth$^4$, Torsten Hoefler$^1$}
\affiliation{
$^1$Dept. of Computer Science, ETH Z\"urich, 8092 Z\"urich, Switzerland \\
$^2$Integrated System Laboratory, ETH Z\"urich, 8092 Z\"urich, Switzerland \\
$^3$IT4Innovations, V\v{S}B - Technical University of Ostrava \\
$^4$Cray UK Ltd. \\
{$^1$\{first.lastname\}@inf.ethz.ch, $^2$\{first.lastname\}@iis.ee.ethz.ch, $^3$jakub.beranek@vsb.cz, $^4$\{sdigirola, droweth\}@cray.com}
}
\renewcommand{\shortauthors}{S. Di Girolamo et al.}

\renewcommand{\paragraph}[1]{{\vspace{0.2em}\noindent\textbf{\textsf{#1}\hspace{0.2em}}}}

\begin{document}

\copyrightyear{2019} 
\acmYear{2019} 
\acmConference[SC '19]{The International Conference for High Performance Computing, Networking, Storage, and Analysis}{November 17--22, 2019}{Denver, CO, USA}
\acmBooktitle{The International Conference for High Performance Computing, Networking, Storage, and Analysis (SC '19), November 17--22, 2019, Denver, CO, USA}
\acmPrice{15.00}
\acmDOI{10.1145/3295500.3356189}
\acmISBN{978-1-4503-6229-0/19/11}

\title{Network-Accelerated Non-Contiguous Memory Transfers}

\begin{abstract}
Applications often communicate data that is non-contiguous in the send- or the
receive-buffer, e.g., when exchanging a column of a matrix stored in row-major
order.
%
While non-contiguous transfers are well supported in HPC (e.g., MPI
derived datatypes), they can still be up to 5x slower than contiguous transfers
of the same size.
%
As we enter the era of network acceleration, we need to
investigate which tasks to offload to the NIC:
%
In this work we argue that non-contiguous memory transfers can be transparently
network-accelerated, truly achieving zero-copy communications. We implement and
extend sPIN, a packet streaming processor, within a Portals 4 NIC SST model,
and evaluate strategies for NIC-offloaded processing of MPI datatypes, ranging
from datatype-specific handlers to general solutions for any MPI datatype.
%
We demonstrate up to 10x speedup in the unpack throughput of real applications,
demonstrating that non-contiguous memory transfers are a first-class candidate
for network acceleration.
\end{abstract}

\settopmatter{printfolios=true}
\maketitle

\section{Motivation}
\enlargethispage{\baselineskip}
The interconnect network is the most critical component of scalable HPC
clusters. After decades of evolution of high-speed networking technologies,
today's HPC networks are highly specialized. This evolution from early
bus-based Ethernet created high-performance switches, OS-bypass, partially
offloaded RDMA networks tuned to move data at hundreds of gigabits/s line-rate
to the application's virtual memory space~\cite{IB}. While all network-specific
packet processing has been moved from the CPU into the Network Interface Card
(NIC) by RDMA, some application-specific processing remains: e.g., sending and
receiving CPUs may need to change the data layout or apply simple computations
(e.g., filtering) to the communication data.  Such data-centric transformations
could be applied while the data is \emph{on the move} through the NIC.

Programmable network acceleration is thus the next logical step to
optimize the interaction between the network and the CPU. The simple
observation that some functions could be efficiently executed on the NIC
has led to various kinds of specialized and limited implementations by
Mellanox~\cite{core-direct}, Cray~\cite{Aries}, and Portals
4~\cite{portals42,schneider-portalsoffload,salvooff}.
However, these acceleration systems can hardly be programmed by the
application developer. Thus, the state of NIC acceleration is similar to
the state of GPU acceleration in the early 2000's where complex shader
languages had to be misused to accelerate computational tasks.
To enable simple application-level network offload, Hoefler et al.
introduced the streaming Processing in the Network (sPIN) concept that
unifies an abstract machine and programming interface for network
accelerators. As such, sPIN is similar to CUDA or OpenCL for compute
accelerators and MPI or UPC for communication.

The sPIN programming model can express arbitrary packet processing tasks
offloaded to the network card from user-level.
Yet, not all tasks are amenable to offload and programmers must
carefully select which pieces of applications to offload to the NIC.
Network accelerated sPIN NICs are designed for data-movement-intensive
workloads and cannot compete with compute accelerators, such as GPUs,
for compute intensive tasks.
The main potential of NIC acceleration arises when the data is
transformed while it moves through the NIC. Here, we expect more
efficient and faster data processing than on a CPU where data items move
through deep memory hierarchies, most likely without any reuse.
Thus, a careful selection of offloaded tasks is imperative for highest
performance.

A class of operations that is amenable to NIC offload is the
transfer of non-contiguous data, an important primitive in many
scientific computing applications. For example, in a distributed graph
traversal such as BFS, the algorithm sends data to all vertices
that are neighbors of vertices in the current frontier on remote
nodes---here both the source and the target data elements are 
scattered at different locations in memory depending on the graph
structure. More regular applications, such as stencil computations in
regular grids used in many PDE/ODE solvers communicate strided data at
the boundaries.
In applications, such as parallel Fast Fourier Transform, the network
can even be used to transpose the matrix on the fly, without additional
copies.
Such non-contiguous data accesses can account for up to 90\% of
application communication
overheads~\cite{mpi-ddt-benchmark,schneider-app-oriented-ping-pong} and
optimizations can lead to speedups of up to 3.8x in
practice~\cite{hoefler-datatypes}.


\enlargethispage{\baselineskip}
Nearly all distributed memory programming and communication interfaces
support the specification of non-contiguous data transfers. Those range
from simple input/output vectors (iovecs) in the standard C library to
recursive derived datatypes in the Message Passing Interface (MPI). With
iovecs, programmers can specify a fixed number of arbitrary offsets in a
message. While this is ideal for implementing communication protocols,
the $\mathcal{O}(m)$ overhead for messages of size $m$ limits its
utility. The second most common interface enables to specify strided
(aka. vector) accesses, typically parameterized by three numbers:
block size, stride, and count. Strided patterns are versatile but limited
to fixed blocks arranged in a fixed stride. The most comprehensive
specification in MPI allows nested types, where a type can be used as a
base-type for another constructor. This specification allows to express
arbitrary data accesses with a single, concise
representation~\cite{gropp-datatype-performance}.

In this work, we explore how to efficiently offload this most complete
and complex specification to practical accelerated network cards. We
study strategies for the implementation and acceleration of arbitrary
derived datatypes.
While many approaches exist for datatype acceleration
(cf.~\cite{schneider-rtcompmpiddt,panda1,panda2}), it is not possible to
transparently offload \emph{arbitrary datatype processing} to network
cards today and data is often received into a buffer and then copied
(``unpacked'') by the CPU. We argue that full offload is a prerequisite
for \emph{true zero copy} communication, leading to significant benefits
in performance and energy consumption.
In short, our main contributions are:
\begin{itemize}[noitemsep,topsep=0pt,parsep=0pt,partopsep=0pt,leftmargin=*]
  \item Design and implementation of \emph{full} non-contiguous memory transfer 
    processing on network
    accelerator architectures.
\item We extend the existing sPIN interface with new scheduling
  strategies to accelerate datatypes processing.
\item We show how sPIN-offloaded DDTs can improve the receiver consumption
bandwidth.
\item We prototype a sPIN hardware implementation that can be integrated into a
NIC in a modern technology node.
\end{itemize}

\section{Background}

Network-acceleration of memory transfers is a critical optimization to improve
application communication phases. Remote Direct Memory Access (RDMA) plays
a major role in this context, allowing the remote processes to perform read
or write operations directly to/from the address space of the target processes.

Non-contiguous memory transfers move data that has to be copied to/from the
target process memory according to a given (non-contiguous) data layout,
making them more challenging to accelerate.
A common solution to implement such transfers is to build
input-output vectors of contiguous memory regions (i.e., memory
offset, size), and offload them to the NIC. However, these
vectors have to be built for each transfer (e.g., different addresses) and are
not space efficient because they always require an overhead linear in
the number of contiguous regions.
In this work we show how next-generation network-acceleration engines, such as
sPIN~\cite{spin}, can efficiently accelerate non-contiguous memory transfers.  We
now provide a brief overview of \spin and discuss the importance of
non-contiguous memory transfers in HPC.

\subsection{Streaming Processing in the Network} \label{sec:spin-background}
\spin is a programming model proposed by Hoefler et al.: it extends the RDMA
and message matching concepts to enable the offloading of simple computing
and data movement tasks to the NIC. Instead of processing full
messages, \spin lets the users specify simple kernels to be executed on a
per-packet basis.  The authors define a Portals 4~\cite{portals42} extension to
support \spin, which we adopt in this paper.

\subsubsection{Portals 4}
Portals 4 is a network programming interface that supports one-sided Remote
Direct Memory Access (RDMA) operations. With Portals 4, a process can expose
parts of its memory over the network by specifying a list entry and appending
it to the Portals 4 priority or overflow lists. The list entries appended to
the overflow list are used as fallback if no list entries are available in the
priority list (e.g., for handling unexpected messages). 

Portals 4 exposes a matching and a non-matching semantic. With the 
matching semantic, a list entry is said \emph{matching list
entry} (ME) and it is associated with a set of match bits.  Whenever a node
wants to issue a remote operation (e.g., put), it specifies the target node on
which the operation has to be executed and the match bits: the operation will
be executed on the area of memory described by a matched ME at the target
(i.e., a ME with the same match bits). The operation is discarded if no
matching MEs are found neither in the priority nor in the overflow list. The interface allows
the matching phase and the operation itself to be executed directly on the NIC. If
the non-matching semantic is used, the operation is executed on the first
list entry available on the priority or overflow list.

Completion notifications (e.g., incoming
operation executed, ack received) are signaled as lightweight counting events or full
events posted on an event queue that can be accessed by the application. 

\begin{figure}[h]
	\vspace{-0.5em}
	\centering{\includegraphics[width=0.9\columnwidth]{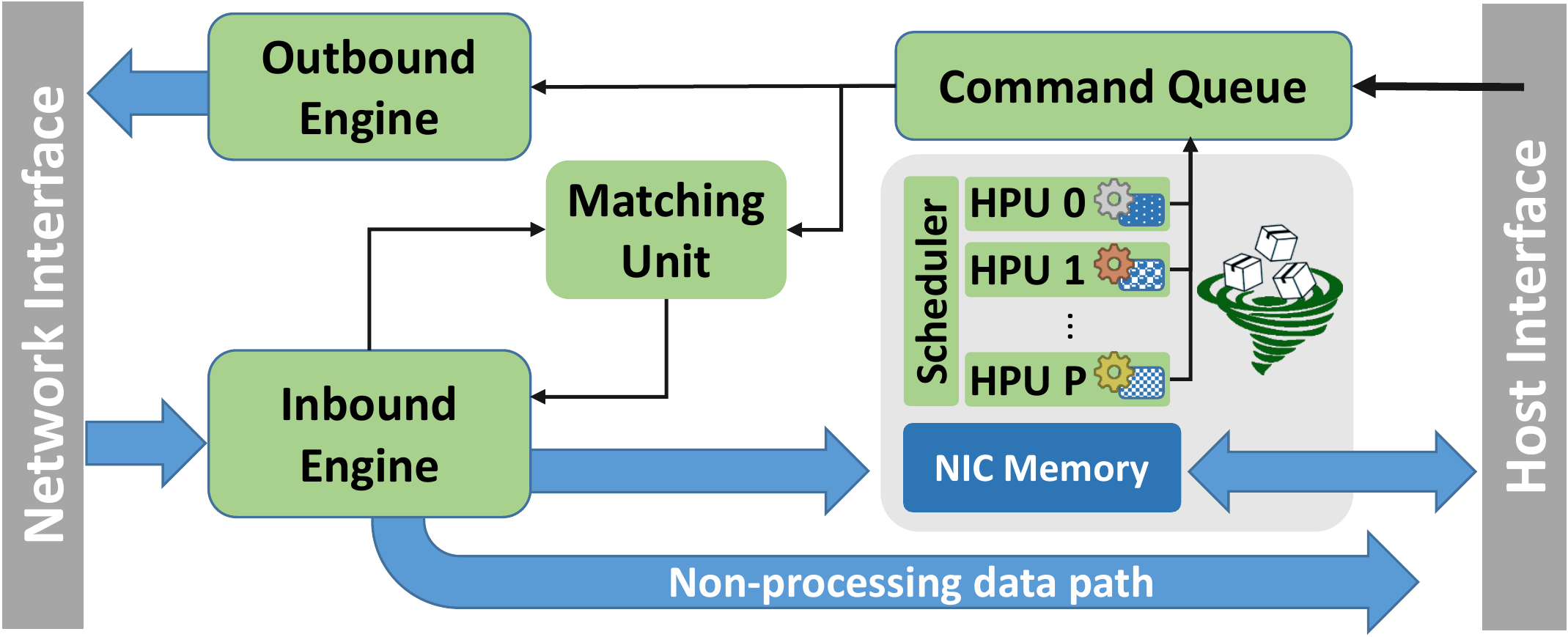}}
	\vspace{-0.5em}
	\caption{sPIN NIC Model}
	\vspace{-2em}
	\label{fig:spin-model}
\end{figure}

\subsubsection{NIC Model}
Fig.~\ref{fig:spin-model} shows a high level view of a NIC implementing \spin.
The message packets are delivered to the \textit{inbound engine} which is
responsible for parsing them and initiates the matching.  There are three packet
types: header (i.e., first packet of the message), completion (i.e., last
packet), and payload (i.e., all the packets between the header and completion
ones).
The matching unit implements the Portals 4 priority and overflow lists: a
message matches a memory descriptor (i.e., matching list entry or ME) if their
match bits are equal.  If the matching
unit receives a matching request for a header packet, then the priority and the
overflow list gets searched according to the Portals 4 matching semantic.
After being matched, an ME can be unlinked from its list but is kept by the
matching unit until the completion packet is received, so to match the rest of
the message packets. We assume that the network delivers the header and
the completion packets as the first and the last ones of a message, respectively.
%

\subsubsection{Executing Handlers}
A ME determines whether a packet has to be processed by \spin.  The
set of handlers to be executed for the header, payload, and completion packets
of a message, together with a NIC memory area that is shared by the handlers,
is defined as \textit{execution context}.
The host application defines the execution context and associates it with an
ME.  If the matched ME is associated with an execution context, then the
packet is copied in the NIC memory and a \textit{Handler Execution Request}
(HER) is sent to the scheduler, which will schedule the handler execution on a 
\textit{Handler Processing Unit} (HPU). If the matched ME has no execution context
associated with it, the data is copied to the host via
the non-processing path.  The packet gets discarded if no matching entries are
found.

\begin{figure}[h]
    \vspace{-1em}
    \centering{\includegraphics[trim=18 0 10 0, clip,
    width=0.95\columnwidth]{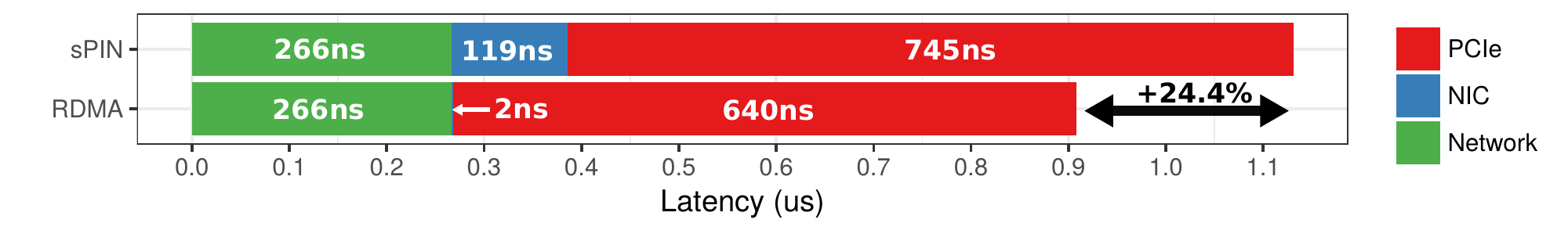}}
    \vspace{-1.5em}
    \caption{Latency of a one-byte put operation.}
    \vspace{-1.5em}
    \label{fig:microbench_latency}
\end{figure}

Fig.~\ref{fig:microbench_latency} shows the latency of a one-byte put operation
(i.e., from when the data leaves the initiator to when it reaches the host
memory).  The data is collected from the simulation environment of
Sec.~\ref{sim}. With \spin, the latency is increased by $\sim24\%$, that is the
overhead for copying the packet to the NIC memory, scheduling the packet
handler, and let it issue a DMA write command. While the 
latency depends on the complexity of the handlers, this data shows
the minimum overhead introduced by \spin.

\enlargethispage{1\baselineskip}
\vspace{-0.2em}
\subsubsection{NIC Commands}
The host can issue commands by pushing them to the NIC command queue (e.g., to
instruct the \textit{outbound engine} to send data).  The HPUs can move data
between the NIC and the host memory by issuing DMA write/read requests. The
write requests can be \textit{fire and forget} operations: the handler is not
required to wait for transfer completion, allowing for a faster release of the
HPU. The HPUs are also interfaced to the NIC command queue, allowing the
handlers to start new communications (e.g., puts) by issuing NIC commands. The
events (e.g., handlers or memory transfer completion, error states) are pushed
to the event queue associated with the Portals 4 Table Entry to which the ME
belongs.

%

\vspace{-0.5em}
\subsection{Non-contiguous Memory Transfers} \label{sec:ncmt}

A key factor in the design of HPC applications is the memory layout they
operate on. The decision of which layout to adopt is driven by many factors,
such as readability of the code, the ability to vectorize, the ability to
exploit locality and hardware prefetching.

\begin{figure}[h]
    \vspace{-0.5em}
    \centering{\includegraphics[width=0.45\columnwidth]{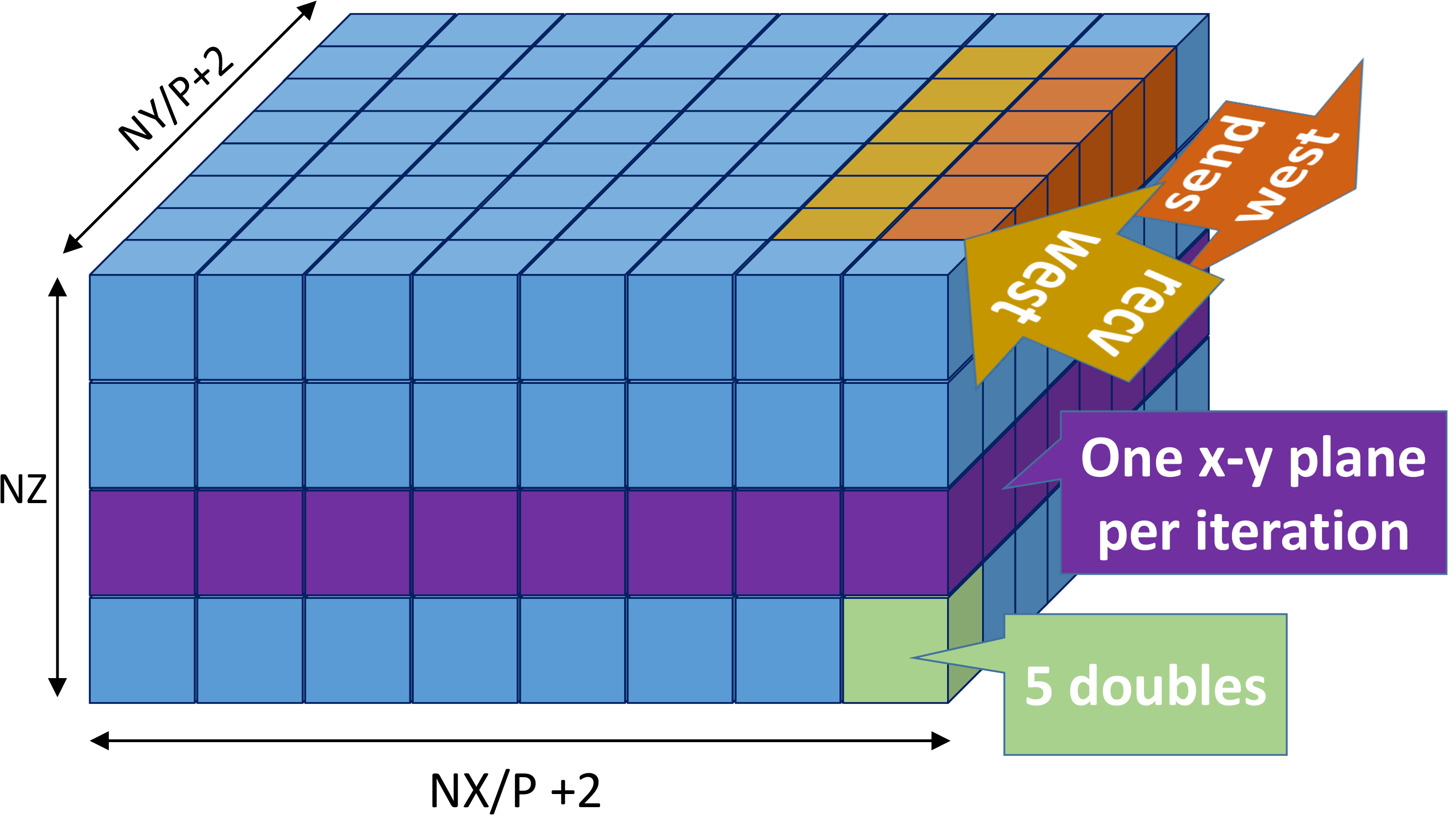}}
    \vspace{-1em}
    \caption{NAS LU data layout and communication.}
    \vspace{-1em}
    \label{fig:naslu}
\end{figure}

As an example, the NAS LU~\cite{van2002parallel, mpi-ddt-benchmark} uses a
four-dimensional array as main data structure. This is split along two
dimensions $(x,y)$ onto a 2D processor grid. The first dimension contains 5
double-precision floats. Each $(x,y)$-plane corresponds to one iteration of the
rhs-solver kernel.  In each communication step, neighbouring faces (i.e.,
$\mathit{nx} \times \mathit{ny} \times 10$ elements) of the four-dimensional
array are exchanged among processors, as sketched in Fig.~\ref{fig:naslu}.


Accelerating \textit{Non-Contiguous Memory Transfers} (NCMT) is a common
problem of 
HPC codes (e.g., the NAS LU benchmark exists since 1993), and (some) support
for non-contiguous communications is present in all major HPC programming
languages. When evaluating non-contiguous memory transfer support, an important
consideration to make is the compactness of the resulting type map: Every
non-contiguous memory transfer can be represented as a mapping from offsets in
the source buffer to offsets in the destination buffer. The goal of
NCMT-support is to exploit regularities in this mapping.
ARMCI~\cite{nieplocha2006high} for example supports two modes to transfer data,
either using a list of source/destination pointers and a block size per
pointer, or strided transfers (multiple stride levels are allowed).
SHMEM~\cite{chapman2010introducing} falls into the same category, however,
strides cannot have multiple levels, thus sending a 3D array slice can require
multiple calls.  Modern HPC languages such as CAF~\cite{mellor2009new},
UPC~\cite{el2006upc}, Chapel and X10 support array slicing within the language
and allow to assign slices to remote arrays but do not support transfers based
on index lists directly (such support might not be needed if the compiler can
aggregate multiple small sends at runtime).

In this paper we chose to focus on MPI~\cite{mpi-3.0} Derived Datatypes, which
allow to specify strided, as well as index-list based transfers and allows
arbitrarily deep nesting of type map descriptions. This decision was made
because MPI supports the widest variety of type constructors, the NCMT support
in any other of the mentioned languages/APIs can be easily mapped to MPI
Derived Datatypes.

\enlargethispage{1\baselineskip}
\vspace{-0.2em}
\subsubsection{MPI Derived Datatypes}
\begin{sloppypar}
The simplest MPI Derived Datatypes (DDTs) directly map to the basic types of
the languages interfaced with MPI, e.g., \texttt{MPI\_INT} maps to an
\texttt{int} in C. We call these predefined DDTs \textit{elementary types}.
In MPI, an integer stored in a variable \texttt{v} can be sent with
\texttt{MPI\_Send(\&v, 1, MPI\_INT, ...)}. Now, if we need to send \texttt{N}
contiguous integers we can either change the \textit{count} argument in the
\texttt{MPI\_Send} call or construct a new DDT
which describes this contiguous data-layout, using
\texttt{MPI\_Type\_contiguous(N, MPI\_INT, \&newtype)}, and supply newtype as
the type of the \texttt{MPI\_Send()}, setting \textit{count} to 1.

To send blocks of data with a regular stride, e.g., a column of an \texttt{N} by
\texttt{N} matrix (stored in row-major layout) we can construct an appropriate
DDT using \texttt{MPI\_Type\_vector(N, 1, N, MPI\_INT, \&newtype)}. The
arguments specify the \textit{count}, \textit{blocklength}, \textit{stride},
\textit{basetype}, \textit{newtype}.  If the data we need to communicate
consists of a mix of types, i.e., an array of structs, we can use
\texttt{MPI\_Type\_create\_struct\-(count, blocklens, displs, types, newtype)}
to construct an MPI DDT that maps to one elment of the array, where
\texttt{count} refers to the number of elements of the struct, \texttt{displs}
gives the displacement for each struct entry in bytes (relative to the address
of the struct itself), \texttt{types} is an array of MPI DDTs, one entry for
each struct member.  If the data we want to send is irregularly strided, we can
use the \texttt{MPI\_Type\_create\_indexed\_block(count, blocklength, displs,
basetype, newtype)} type constructor, which allows to pass a list of
displacements and length per block.
MPI supports additional type constructors not discussed here (e.g.,
\texttt{MPI\_Type\_create\_subarray()}).  Some DDT constructor calls
described above support variants to, e.g., specify the displacements in bytes instead
of multiples of \textit{basetype}.

All MPI DDTs need to be \textit{committed} before they can be used in any
communication call.  An MPI implementation which aims to optimize
the implementation of DDTs can intercept the commit call to e.g.,
runtime-compile DDTs or prepare for their network offload.

Data can be locally packed and unpacked outside the communication functions
with the \texttt{MPI\_Pack()} and \texttt{MPI\_Unpack()} calls, respectively. 
However, in doing so we forgo all possible optimizations MPI could perform,
such as zero-copy data transfer, pipelining packing/sending, etc. The same is
true for codes which perform ``manual packing'' of data before sending, a
practice widespread in old HPC codes, due to the fact that a compiler can
optimize the packing loops of specialized codes, e.g., utilize vector
instructions to copy blocks, while a simple MPI implementation might only
provide a non-specialized generic DDT interpreter.  However, much effort has
been directed into optimizing MPI DDT implementations, and using them often
gives superior
performance~\cite{gropp1999improving,byna2006automatic,tanabe2008introduction}.
\end{sloppypar}

\section{Accelerating Non-Contiguous Memory Transfers}

\begin{figure}[h]
    \vspace{-0.5em}
    \centering{\includegraphics[width=0.9\columnwidth]{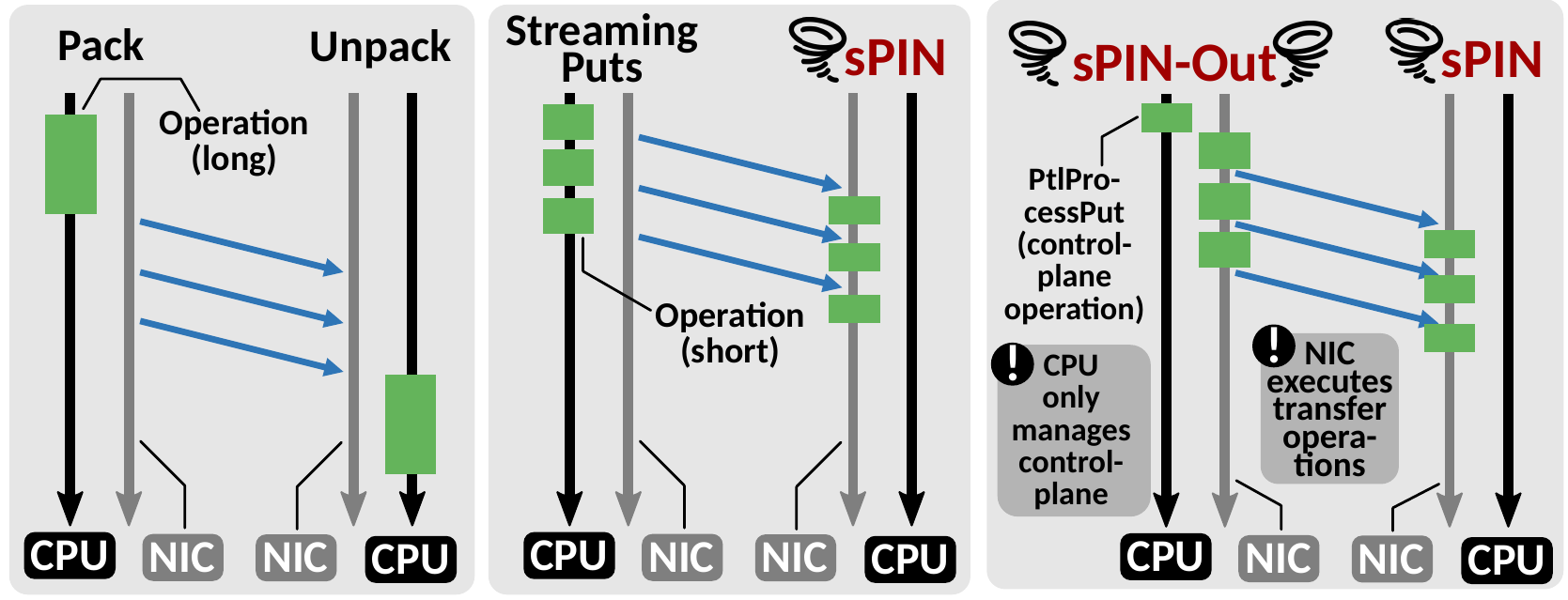}}
    \vspace{-1em}
    \caption{Network-accelerated DDT processing strategies.}
    \vspace{-2em}
    \label{fig:ddtoff_overview}
\end{figure}

%
Fig.~\ref{fig:ddtoff_overview} sketches three possible implementations of
non-contiguous memory transfers. The sender and receiver side implementations
can be interchanged creating different solutions.
%
%
The left tile shows the non-accelerated case: the sender CPU packs the
data in a contiguous buffer before sending it and the receiver CPU unpacks it
after receiving it. 
While this approach is efficient for small message sizes, the
packing and unpacking overheads can limit the transfer performance of large
messages.
In the middle tile, the sender avoids the packing phase by streaming contiguous
regions of data as they are identified. The receiver processes the incoming
packets directly with \spin: for each packet, a handler
identifies one or more contiguous regions where the payload
has to be copied to.
However, the sender CPU is still busy finding the contiguous regions. This
phase can be fully overlapped if we put \spin on the NIC outbound path (right
tile): the put-activated handlers can take care of the sender datatype
processing and the issuing of the data transfers.

\subsection{Sending Non-Contiguous Data}
In this section we outline the different strategies to accelerate
datatype processing at the sender side.

%

\subsubsection{Streaming Puts}
The packing phase can be optimized by sending contiguous regions of data as
they are identified, overlapping this communication with the search of the next
region. Practically, this means splitting the send of the packed data in
multiple put operations. However, the data sent by these multiple puts need to
be seen as a single message by the receiver to, i.e., allow \spin enforcing
schedule dependencies and minimize the resource usage (e.g., events, match list
entries).

We extend the Portals 4 interface introducing \textit{streaming puts}, that allow
to specify the message data via multiple function calls.
A streaming put is started with a \texttt{PtlSPutStart} call,
which is similar to a \texttt{PtlPut} but takes an additional parameter that is
the contiguous memory descriptor $\langle
\text{offset}, \text{size} \rangle$ identifying the data to be sent by this call.
The applications can keep sending data as part of the same put operation by
using \texttt{PtlSPutStream} calls, that take as arguments: the streaming put
descriptor, a contiguous memory region descriptor, and an
\textit{end-of-message} flag to specify if this call concludes the put
operation.
While Portals 4 already allows the same ME to be matched by different puts,
these are seen as different messages, each of them triggering the message
matching process, being processed as different messages by \spin, and
generating different events. A streaming put is transparent to the target since
the packets generated by the different streaming put calls are part of
the same message.

\subsubsection{Outbound sPIN}
Streaming puts improve the sender-side datatype processing performance but
still require the sender CPU to be involved in the process.
Full offloading of the sender-side datatype processing can be achieved by
extending sPIN with the ability to process also packets that are being
sent out. 
The idea is to have a new type of put operation (namely \texttt{PtlProcessPut})
that signals the NIC to create the packets of the message but, instead
of injecting them on the network, forward them to the internal packet
processing unit. In particular, the outbound engine creates a Handler Execution
Request for each packet and sends it to the packet scheduler (as it happens for
the incoming packets). 
In this case, we do not have the urgency to buffer the packet data in NIC
memory since this is already stored in the host memory: for a
\texttt{PtlProcessPut}, the outbound engine does not fill the packet with data (i.e.,
by DMAing it from host memory) but
delegates this task to the packet handler. 
%

In this model, the sender specifies an execution context to associate with the
\texttt{PtlProcessPut} operation. This put operation leads to the generation of
the packets and, for each one of these, a handler is executed on the sender
NIC. The handler is in charge to identify the contiguous area of memory that
the packet needs to carry (similarly to receiver-side datatype processing, see 
Sec.~\ref{sec:recv_data}) and send the data out using a streaming put. 
%

%



\subsection{Receiving Non-Contiguous Data}\label{sec:recv_data}
We now describe how incoming messages are processed in \spin
according to a given datatype.
We first introduce \textit{packet scheduling policies}, an extension to \spin
that enables the user to influence the handler scheduling on the HPUs for a
given message, then we describe the \spin execution context
to process datatypes on the NIC. Finally, we discuss two different approaches
to datatype processing: with specialized or general handlers.

\subsubsection{Packet Scheduling Policies}\label{sec:spin-scheduling}
The \spin scheduler enforces happens-before dependencies on the
handlers execution: i.e., for a given message, the header handler
executes before any payload handler and the completion handler executes after
all the payload handlers.
Ready-to-execute handlers are assigned to idle HPUs.
%

We extend the \spin execution context to let the user specify scheduling
policies different than the default one and implement a blocked round-robin
packet scheduling policy, namely \textit{blocked-RR}.
With \textit{blocked-RR}, sequences of $\Delta p$ consecutive packets are
sequentially processed: no two HPUs can execute handlers on the packets of
this sequence at the same time.
To avoid keeping HPUs busy waiting for the next packet of a sequence,
we introduce \textit{virtual HPU}s (vHPUs). They now become our
scheduling units: a sequence of packets is assigned
to a vHPU, that is scheduled to run on a physical HPU.
The vHPU is in charge to process the handlers of the packets in the assigned
sequence: if there are no packets to execute, the vHPU yields the HPU and gets
rescheduled whenever a new packet of that sequence arrives. Packets of the same
sequence can be processed out-of-order by the vHPU. The number of vHPUs and
$\Delta p$ are parameters of the scheduling policy.

\subsubsection{DDT-Processing Execution Context}

To offload datatype processing we need to create an ME to receive the data
and a \spin execution context to process it.
In particular, we need to (1) install the header, payload, and completion handlers; (2)
allocate and initialize the NIC memory needed for the datatype processing; and
(3) define the packet scheduling policy associated with this execution context.
We now describe the general structure of the handlers, deferring their detailed
implementation, the NIC memory management, and the
scheduling policy selection to the following sections.

\begin{itemize}[noitemsep,topsep=0pt,parsep=0pt,partopsep=0pt,leftmargin=*]

\item \textbf{Header handler.} For datatype processing we do not perform any
operation in the header handler, hence this is not installed.

\item \textbf{Payload handler.} The payload handler identifies all the contiguous
regions in receiver memory that are contained in a packet. For each one of them,
it issues DMA write requests towards the receiver memory.
Blocking and non-blocking DMA write requests are defined in \spin.  We always
use non-blocking calls (i.e., \texttt{PltHandlerDMAToHostNB}) because our
payload handlers do not need to wait for transfers completion.  We extend this
call adding the option \texttt{NO\_EVENT} that, when specified, avoids that the
transfer completion generates an event on the host.

\item \textbf{Completion handler.} The completion handler issues a final
zero-byte \texttt{PltHandlerDMAToHostNB} without the \texttt{NO\_EVENT}
option, so to generate the transfer completion event and signal the host that
all the data has been unpacked.

\end{itemize}

\subsubsection{Specialized Payload Handlers} \label{sec:specialized_handlers}

To process incoming packets and derive the destination memory offsets, the
payload handler needs to be aware of the datatype describing the memory layout.
This awareness can be achieved with datatype-specific handlers or
with generic ones operating on a datatype description.
We now describe a set of handlers that are specialized for MPI Derived
Datatypes having as base types an elementary type (e.g., \texttt{MPI\_INT},
\texttt{MPI\_FLOAT}) or contiguous types of elementary types.
It is worth noting that in some cases more complex (i.e., nested) datatypes can
be transformed to simpler ones via datatype
normalization~\cite{Traff:2014:OMD:2642769.2642771}, 
potentially making them compatible with the specialized handlers.

%
The MPI \textit{vector} datatype identifies a non-contiguous memory region that
is composed of a fixed \textit{number of blocks}. Blocks are composed of a
\textit{number of elements} of a certain \textit{base type} and start
\textit{stride} elements apart from the previous one.
To offload this datatype processing to \spin, the host allocates a
\texttt{spin\_vec\_t} data structure in the NIC memory containing the
parameters described above.
The pseudocode of the vector-specialized payload handler is shown in
Listing~\ref{lst:vec}. To improve readability, we show a simplified version of
the handler, that does not handle the corner case where the block size does not
divide the packet payload size.
The handler first determines the host address where the first block of the
packet has to be written to, then it proceeds to copy all the blocks contained
in the packet payload at the right offsets (i.e., every \textit{stride} bytes).

\vspace{0.5em}
\begin{lstlisting}[caption=Vector-specialized payload handler pseudocode, label=lst:vec]
int vector_payload_handler(handler_args_t *args){
  uint8_t    *pkt_payload   = args->pkt_payload_ptr;
  uint8_t    *host_base_ptr = args->ME->host_address;
  spin_vec_t *ddt_descr     = (spin_vec_t *) args->mem;
  uint32_t   block_size     = ddt_descr->block_size;
  uint32_t   stride         = ddt_descr->stride;
  uint32_t host_offset = (args->pkt_offset / block_size)*stride;
  uint8_t    *host_address  = host_base_addr + host_offset;

  for all the blocks in the packet payload {
    DMA write (pkt_payload, block_size) to host_address
    pkt_payload += block_size; host_address += stride;
  }
  return PTL_SUCCESS; }
\end{lstlisting}

\paragraph{Other datatypes.} More irregular memory areas can be described with
more complex datatypes.  For example, the \textit{index-block} datatype models
fixed-size blocks of elements displaced at arbitrary offsets.
The \textit{index} and \textit{struct} datatypes extend it allowing
to model variable size, or size and base type, memory regions, respectively.

To process these datatypes in \spin, we need to provide the handlers
with additional information (e.g., list of offsets and block sizes).
The host application is in charge to copy this information to the NIC memory.
The datatype influences also the handler complexity: e.g., to find
the correct offset and block size for a given packet (or portion of a packet)
we use a modified binary search on these lists that have size linear in
the number of non-contiguous regions.

\subsubsection{General Payload Handlers}\label{sec:general_handlers}

Having specialized handlers always guarantees the best performance, but it is
impossible to provide them for any arbitrary derived datatype.  While the users are
free to write handlers for their own datatypes (e.g., similar to writing custom
unpack function), this solution does not apply to the general case.
In this section we discuss a set of approaches to transparently process any
derived datatype in \spin, without the need of ad-hoc handlers.
Our general handlers are based on the \textit{MPITypes}
library~\cite{ross2009processing}, that enables partial processing of
MPI datatypes outside MPI.
We optimize the library for faster offloaded execution by removing MPI
dependencies and exchanging memory copy instructions with DMA write requests.

\begin{figure}[h]
    \vspace{-0.6em}
    \centering{\includegraphics[trim=0 0 0 0, clip,
    width=\columnwidth]{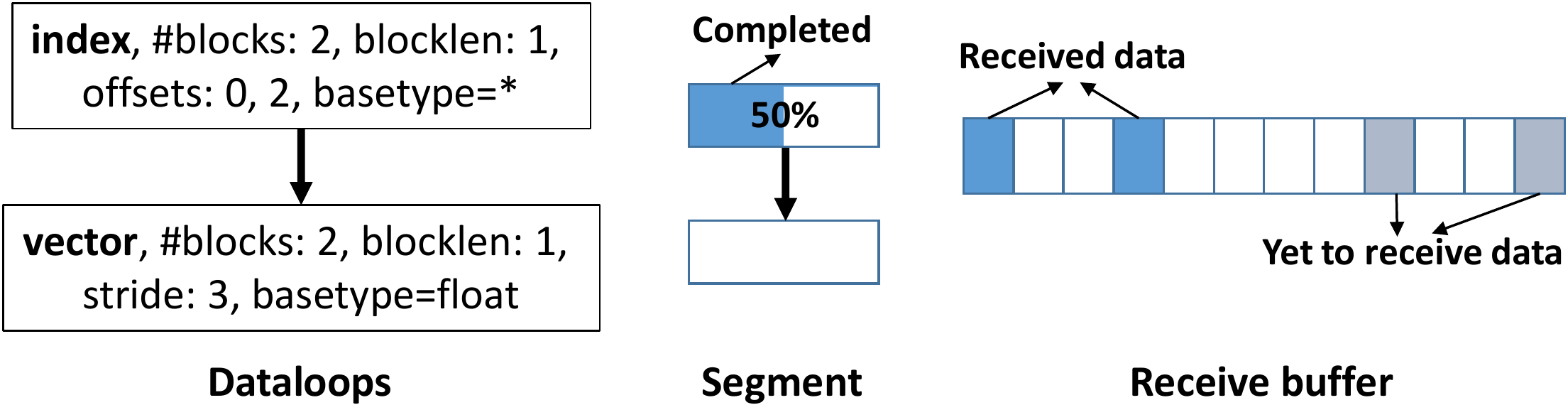}}
    \vspace{-2em}
    \caption{MPITypes dataloops for an index of vectors datatype; the
    segment representing the datatype processing state; and the receive buffer
    where the data is unpacked.}
    \vspace{-1em}
    \label{fig:mpitypes_example}
\end{figure}

\paragraph{MPITypes.}
MPITypes library represents datatypes as sets of descriptors called
\textit{dataloops}. There are five types of dataloops: contig, vector,
blockindexed, indexed, and struct~\cite{ross2003implementing}.
Dataloops having elementary types as base type are
defined as \textit{leaves}. The processing of a datatype is split in two
phases: (1) the non-leaf dataloops are traversed in order to determine the
offset of the leaf, (2) the leaf dataloop is processed to determine
the relative offset in the receive buffer.
Leaves are processed with specialized functions, similar to the specialized
handlers of Sec.~\ref{sec:specialized_handlers}.

The partial progressing of datatypes is achieved by exporting the datatype
processing state to a structure called \textit{segment}. The state is
represented by a stack, modeling the recursive processing of dataloops, and each
element of the stack is a dataloop state.
Fig.~\ref{fig:mpitypes_example} shows an index of vectors datatype expressed with
dataloops (left). The segment state (middle) shows that the index
dataloop has already been progressed by $50\%$: i.e., one block (which is a vector
datatype) has already been unpacked in the receive buffer (right).

The packed data is modeled as a stream of bytes: the datatype processing
function takes a \textit{first} and \textit{last} bytes as input, representing
the portion of stream to be processed (we always process one packet payload at a time).
If the specified \textit{first} byte is after the last byte processed in a
given segment, a \textit{catch-up} phase takes place: the segment is progressed
(without issuing DMA writes) until the position \textit{first} is reached.
Instead, if the \textit{first} byte is before the last processed byte, then the
segment is reset to its initial state.

The datatype processing function advances the state of the segment passed
as input. In \spin, multiple packets of the same message can be processed
at the same time by different HPUs, generating write conflicts if all
of them use the same MPITypes segment.
While a possible solution is to enforce mutual exclusion on the critical
section inside MPITypes, this adds synchronization overheads that can serialize
the handlers execution.
We now discuss three approaches to datatype processing offload with MPITypes
that avoid write-conflicts without requiring synchronization.

\paragraph{HPU-local.}
This strategy replicates the MPITypes segment on each vHPU, using blocked-RR as
packet scheduling policy with $\Delta p=1$ and the number of vHPUs $P$ equal to
the number of physical ones.  While this solution completely removes the
write-conflicts problem (i.e., each vHPU writes to its own segment), it is
characterized by long catch-up phases during the datatype processing.  In fact,
assuming in-order packet arrival, each vHPU gets every other $P-1$ packet: for
each packet, it has to progress its state by $P-1$ old packets (i.e., catch-up
phase) before getting to the assigned one.
In case of out-of-order packet delivery, a vHPU may receive a packet that is
before the one processed last in its segment. In this case, the vHPU-local
state is reset, incurring an additional overhead.

\begin{figure}[h]
    \vspace{-0.5em}
    \centering{\includegraphics[trim=0 0 0 0, clip,
    width=0.8\columnwidth]{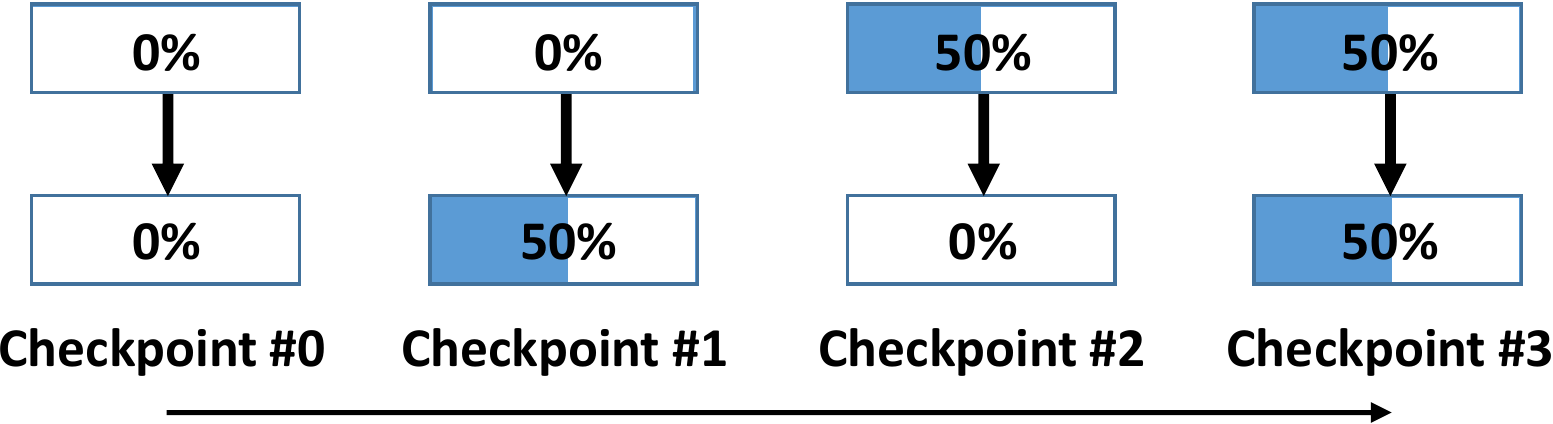}}
    \vspace{-1em}
    \caption{Checkpointing the segment state.}
    \label{fig:mpitypes_checkpoint}
    \vspace{-1em}
\end{figure}

\paragraph{RO-CP: Read-Only Checkpoints.}
The idea of RO-CP is to solve the write-conflicts problem by avoiding the
handlers to write back to shared memory.  This is possible if each handler
makes a local copy of the segment and then start processing on it, without
writing the modified copy back.  However, this would require each handler to
always start the processing from the initial segment state.  
To address this
issue, we introduce \textbf{checkpoints}: \textit{a checkpoint is a snapshot of
the MPITypes segment processing state}. Fig.~\ref{fig:mpitypes_checkpoint}
shows different checkpoints of the same segment: the datatype is processed on
the host and every $\Delta r$ bytes (we define $\Delta r$ as checkpoint
interval) a copy of the segment is made and used as checkpoint.

Whenever a handler executes on a packet, it selects the closest checkpoint that
can be used to process the packet payload, makes a local copy of it, and then
starts processing.
This solution bounds the handler runtime to $O(\Delta r)$, but introduces the overhead
of creating the checkpoints and copying them to the NIC.
The checkpoint interval enables a tradeoff between the payload handler runtime
and the NIC memory space needed to store the checkpoints: i.e., the smaller the
checkpoint interval, the faster the handlers but also the more the checkpoints.
For $\Delta r=k$, where $k$ is the packet size, i.e., one
checkpoint per packet: no catch-up phase and local segment copy are required
(the checkpoint is used only once) but $\left \lceil \frac{m}{k} \right \rceil$
checkpoints are stored in  NIC memory.

\paragraph{RW-CP: Progressing Checkpoints.}
The performance of the HPU-local and RO-CP is undermined by the long catch-up phase
and the checkpoint copy. We now discuss a solution that avoids the catch-up
phase without requiring additional checkpoint copies in case of in-order
packet arrival.
The idea is to exploit the performance optimal case of RO-CP (i.e.,
\mbox{$\Delta r = k$}), assigning sequences of $\Delta r$ consecutive packets
to the each vHPU, that can now act as exclusive owners of a checkpoint. In this
way we avoid the catch-up phase and the local copy even when $\Delta r > k$.
This strategy can be implemented by using the blocked-RR scheduling policy 
(see Sec.~\ref{sec:spin-scheduling}) with
$\Delta p = \left \lceil \frac{\Delta r}{k} \right \rceil$ and a number of vHPU
equal to the number of packet sequences: all the packets using the same checkpoint
get assigned to the same vHPU.
In case of out-of-order packet delivery, the checkpoint state may be ahead of
the received packet offset: in this case the changes to the checkpoint need
to be reverted. 
We keep a master copy of the checkpoints in NIC memory, so they can be
reset to their original checkpointed state if needed.

\begin{figure}[h]
    \vspace{-0.5em}
    \centering{\includegraphics[trim=0 0 0 0, clip,
    width=0.9\columnwidth]{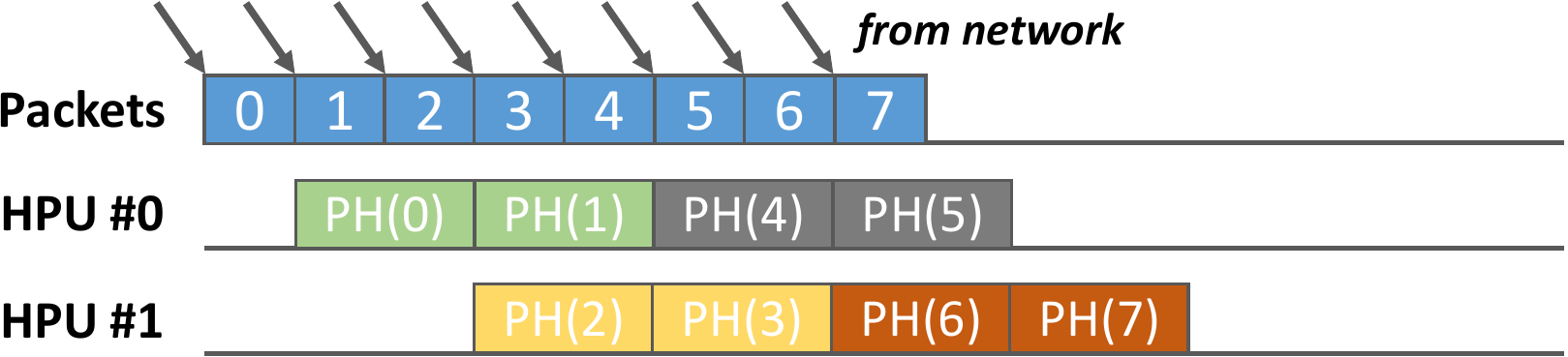}}
    \vspace{-0.9em}
    \caption{Packets scheduling in RW-CP ($\Delta r = 2 \cdot k$).}
    \label{fig:rwcp-scheduling}
    \vspace{-1em}
\end{figure}

\paragraph{How do we select the checkpoint interval?}
While RW-CP optimizes the handler completion time, the employed blocked-RR
scheduling policy introduces a packet scheduling dependency (i.e., the packets
belonging to the same sequence cannot be processed in parallel, see
Fig.~\ref{fig:rwcp-scheduling}), that influences the whole message processing
time.
We define message processing time as time from when the first byte of a
message is received to when the last byte of the message is copied to the
receiver buffer. We now model the message processing time for the RW-CP
strategy, and discuss how to limit the scheduling overhead without overflowing
the NIC memory.

Let us define the effective packet arrival time as $T_{pkt}$, the total number
of packets composing a message of size $m$ bytes as \mbox{$n_{pkt} = \left \lceil
\frac{m}{k} \right \rceil$}, and $P$ as the number of physical HPUs. We assume
that $P < n_{pkt}$ for the sake of simplicity.
The general payload handler runtime can be modeled as:
\[ T_{PH}(\gamma) = T_{PH}^{init} + T_{PH}^{setup} + \gamma \cdot T_{PH}^{block} \]
where $\gamma$ is the average number of contiguous memory blocks per packet,
$T_{PH}^{init}$ is the time needed to start the handler and prepare the
arguments to pass to MPITypes (e.g.,
copy of the segment in RO-CP), $T_{PH}^{setup}$ is the startup overhead of the
datatype processing function (e.g., includes the catch-up phase),
$T_{PH}^{block}$ is the time needed by the datatype processing function to find
the next contiguous block.

The message processing time of RW-CP can be modeled as:
\[ T_{C} = T_{pkt} + \left \lceil \frac{\Delta r}{k} \right \rceil \cdot (P-1) \cdot T_{pkt} +
\left \lceil \frac{n_{pkt}}{P} \right \rceil \cdot T_{PH}(\gamma) \]
that is the time needed to receive/parse the first packet ($T_{pkt}$), plus
the time to saturate the HPUs (we can schedule a new vHPU every 
$\left \lceil \frac{\Delta r}{k} \right \rceil$ received packets), plus the time 
taken by the last
vHPU to process its share of packets (each vHPU gets 
$\left \lceil n_{pkt} / P \right \rceil$ packets).

The message processing time gets minimized for $\Delta r = 1$, but this
requires to store a number of checkpoints equal to the number of packets.  We
observe that the overhead induced by $\Delta r$ gets negligible for large
message sizes or for large values of $\gamma$. Hence, we can
define a strategy to compute a checkpoint interval such that:
\begin{itemize}[noitemsep,topsep=0pt,parsep=0pt,partopsep=0pt,leftmargin=*]
\item The scheduling dependency overhead is less than a factor $\epsilon$ of the packets
processing time:
\[ T_{pkt} + \left \lceil \frac{\Delta r}{k} \right \rceil \cdot (P-1) \cdot T_{pkt} \le \epsilon
\cdot \left \lceil \frac{n_{pkt}}{P} \right \rceil \cdot T_{PH}(\gamma) \]
\item The total number of checkpoints fits in the NIC memory:
\[ \frac{n_{pkt} \cdot k}{\Delta r} \cdot C \le M_{NIC} \]
where $C$ is the checkpoint size (612\,B in our configuration).
\item The packets buffered during the scheduling policy overhead
fit in the packet buffer ($B_{pkt}$ is the packet buffer size in bytes):
\[ \min \left( \frac{T_{PH}(\gamma) \cdot k}{T_{pkt}}, \Delta r \right) \le B_{pkt} \]
\end{itemize}

\subsubsection{General vs. Specialized Payload Handlers}
\begin{figure}[h]
    \vspace{-0.5em}
    \centering{\includegraphics[trim=0 0 0 0, clip,
    width=1\columnwidth]{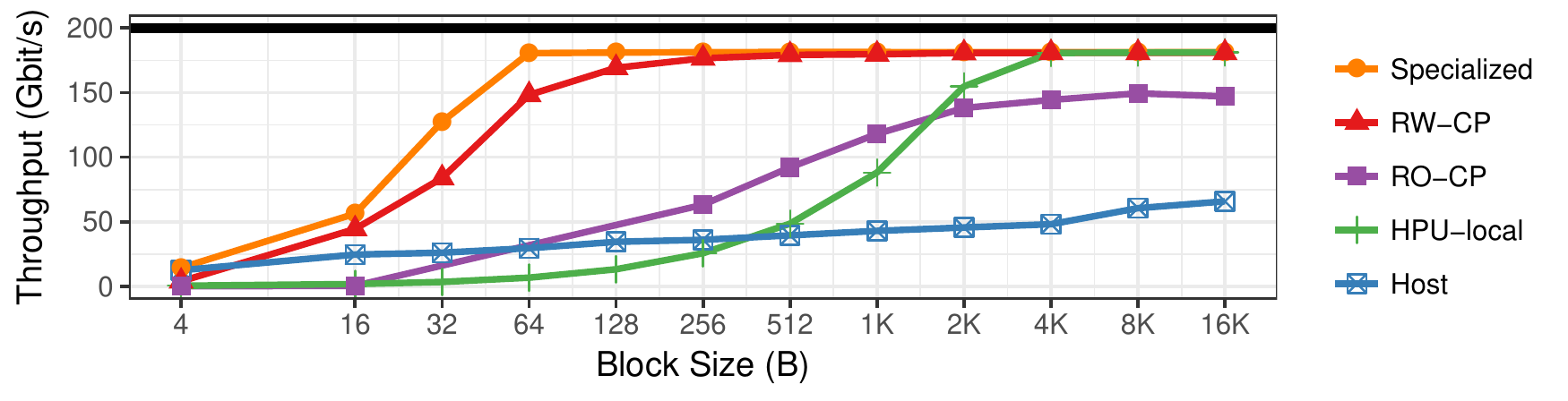}}
    \vspace{-2.6em}
    \caption{Unpacking throughput of an \texttt{MPI\_Type\_vector} as function
    of the block size.}
    \vspace{-1em}
    \label{fig:microbench_vec_block}
\end{figure}

Fig.~\ref{fig:microbench_vec_block} shows the message processing throughput for
a 4\,MiB message with a vector datatype. We vary block size (x-axis) and stride
(twice the blocksize).  The benchmark reports the simulation results for an
ARM-based ($16$ Cortex A15 HPUs @800MHz) sPIN implementation (see
Sec.~\ref{sec:sim-testbench}).

\noindent The specialized handler reaches line rate already for 64\,B block size (i.e.,
$\gamma = 32$). The throughput of RO-CP is limited by the segment copy, while
the catch-up phase of HPU-local shrinks with the block size. The handlers
become slower as the block size decreases (more iterations per handlers), making
offloaded datatype processing slower than host-based unpack (i.e.,  RDMA + CPU unpack)
for 4\,B blocks.


\vspace{-0.2em}
\subsubsection{Integration in MPI}

Before starting to process a message with a given datatype, the host needs to
configure the NIC to make it ready to execute the handlers as the packets
arrive. In this section we discuss how to implement offloaded datatype
processing in communication libraries, using MPI as driving example.

\noindent\textbf{(1) Define the MPI derived datatype} and commit it. During the
commit, the implementation determines the processing strategy to use (e.g.,
specialized or general handlers). according to the datatype being commited  At
this stage, the DDT data structures to offload to the NIC are created: e.g.,
the checkpoints for the RO-CP and RW-CP. 

\noindent\textbf{(2) Post the receive operation.} During the posting of a receive using a
derived datatype, the host tries to allocate NIC memory to store the DDT
data structures. If the allocation fails due to the lack of space, the MPI
implementation is free to (a) fall back to the non-offloaded DDT
processing; or (b) free previously allocated memory (e.g., by applying a LRU
policy on the offloaded datatypes). Once the NIC memory has been allocated and
the DDT data structures copied to it, a ME is created and appended to the
priority list.

\noindent\textbf{(3) Complete the receive operation.}
The MPI implementation gets an event whenever the datatype processing
is over (i.e., all the DMA writes completed) and can conclude the receive.

The user can influence the NIC datatype processing by using the 
\texttt{MPI\_Type\_set\_attr} call to specify the type
attributes. Possible indications are: if the type has to be offloaded or not;
the priority with which this type has to be offloaded (e.g., to drive the
victim selection scheme if no NIC memory space is available during (2)); the
$\epsilon$ parameter discussed in Sec.\ref{sec:general_handlers}.

Offloaded datatype processing is not possible for unexpected messages because
the receiver-side datatype is not known at that stage (a matching receive has
not been posted yet). They can be unpacked by falling back to the host
CPU-based unpack methods.



\begin{figure*}[t]
  \newcommand{\figheight}{.38\columnwidth}
    \includegraphics[height=\figheight]{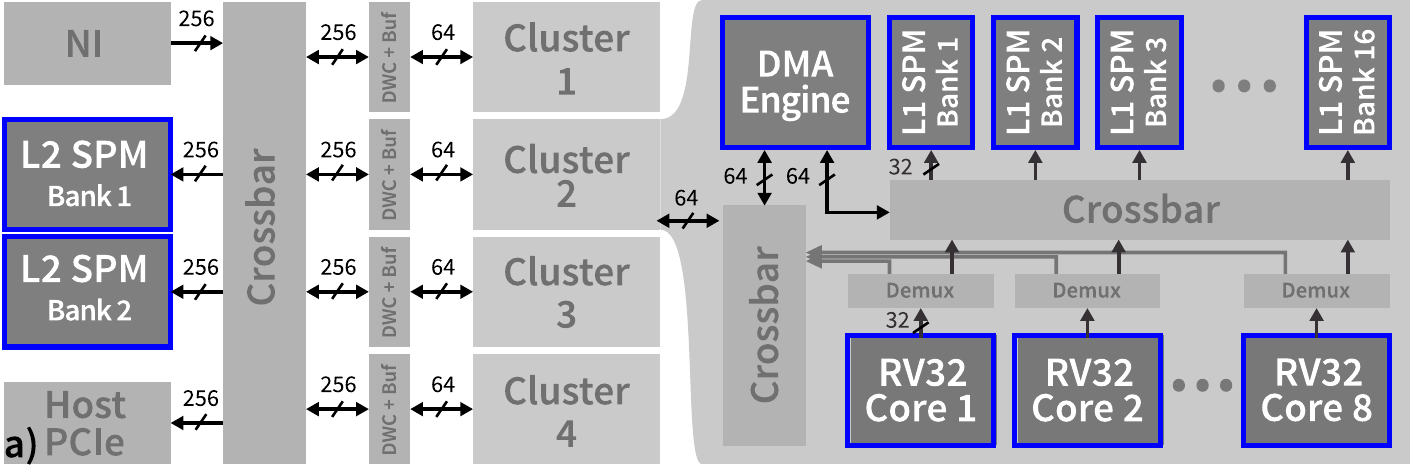}
  \hfill
    \includegraphics[height=\figheight]{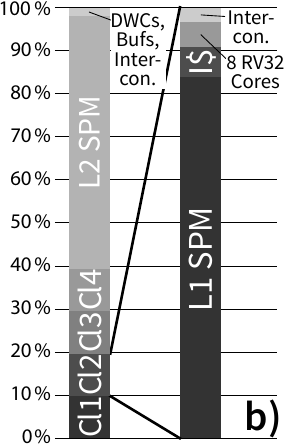}
  \hfill
    \includegraphics[height=\figheight]{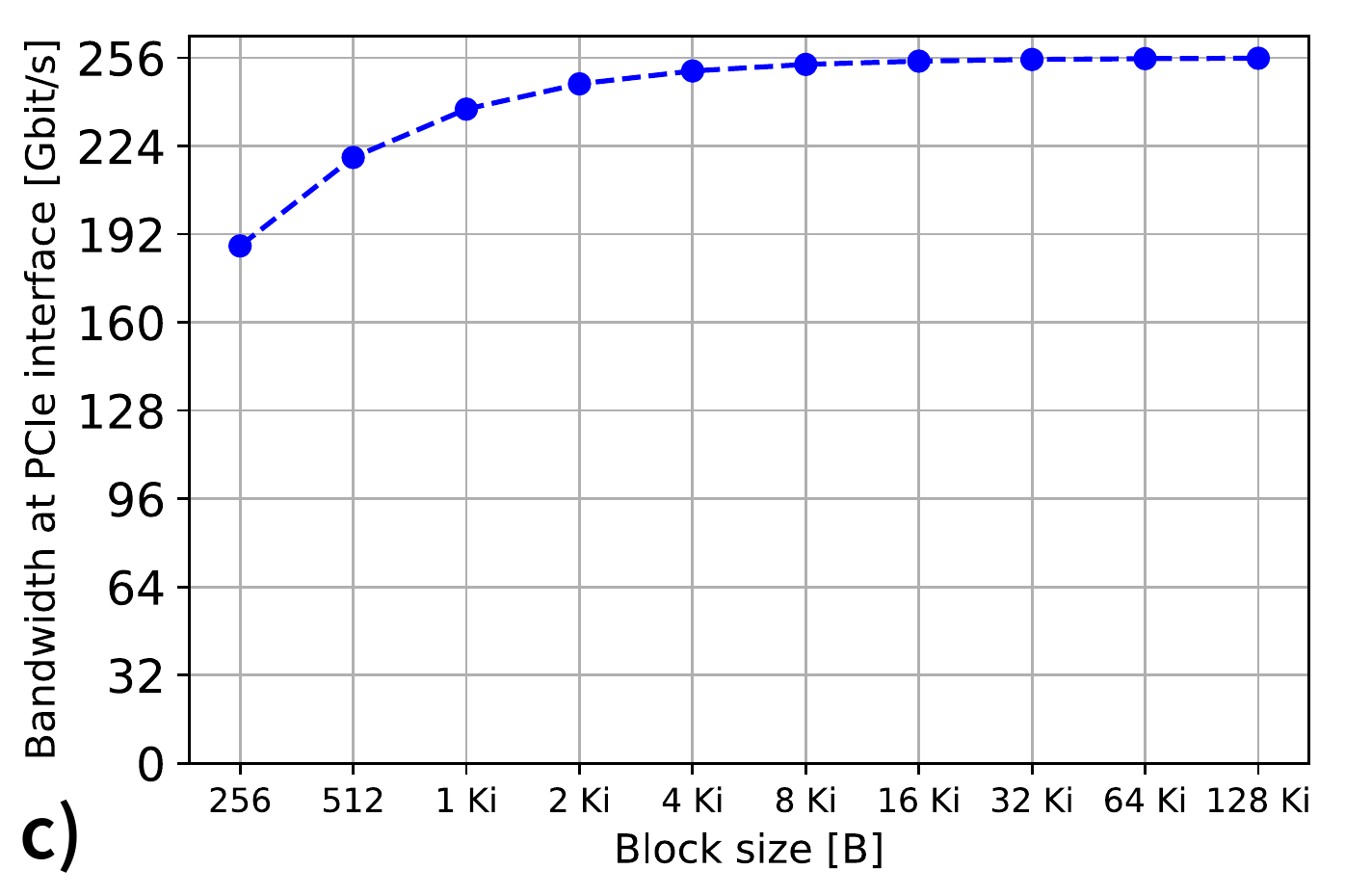}
    \vspace{-1em}
  \caption{a) Proposed hardware architecture of the sPIN accelerator. Single-headed arrows indicate half-duplex, and double-headed arrows full-duplex bandwidth. b) Area breakdown of the accelerator. c) Bandwidth in function of block size.}
    \vspace{-0.5em}
  \label{fig:hw}
\end{figure*}
%

\section{Prototyping {sPIN}} \label{sec:hw}
In this section, we propose a first design of a hardware implementation of 
\spin, presenting first estimates for the complexity and power efficiency for 
a modern 22\,nm technology implementation. Further, we validate the basic performance
requirements in terms of throughput and memory bandwidth by running
micro-benchmarks on a cycle-accurate simulation. 

\paragraph{Design Goals and Requirements.} 
The overall design goal is to fit the \spin accelerator onto the same silicon
die as the NIC, while ensuring that it is capable of sustaining a line rate of
200\,Gbit/s. Hence the accelerator memory bandwidth must be matched, and the
processing power must be aligned with the expected computational load. On
typical DDT benchmarks we measured an average operational intensity (OI) around
0.04 - 0.08\,op/B (w.r.t. the NIC input bandwidth), which turns into a load of
1-2\,Gop/s at 200\,Gbit/s. The use cases considered can allocate up to 3\,MiB
of temporary storage and hence a local memory with more than 6\,MiB should be
employed to enable efficient operation with double-buffering.

\subsection{Hardware Architecture Overview}
Since an area and energy-efficient microarchitecture is crucial to make
co-integration of \spin accelerator and NIC feasible, we leverage the parallel
ultra low power (PULP) platform for this
estimation~\cite{kurth2017hero,Rossi2017}. PULP is a clustered
many-core architecture that employs fully programmable 32\,bit RISC-V
cores~\cite{gautschi2017ri5cy} as main processing elements. The PULP
architecture is geared towards area and energy-efficiency, and hence data
movement is controlled in software using DMA transfers between local scratchpad
memories (SPMs) in order to minimize communication-related overheads. Apart
from being more efficient than designs with hardware-managed
caches~\cite{Rossi2017}, this approach ensures tight control over the compute
and data movement schedule, leading to predictable execution patterns with low
variance. Hence, PULP is an ideal architectural template for \spin,
and the fact that the project is completely open-source
facilitates the analysis and dissemination of the results.

Fig.~\ref{fig:hw}a shows a high-level overview of the proposed accelerator
architecture, which is based on a PULP multicluster~\cite{kurth2017hero} that
has been modified to meet the requirements of this application (the central
elements are highlighted with a blue border). The analyzed configuration
comprises four clusters with eight cores each. To stay above the memory
requirements, the architecture contains 12\,MiB of memory distributed over two
hierarchy levels with 16$\cdot$64\,KiB L1 SPM banks per cluster and
{2$\cdot$4\,MiB L2 SPM banks on the top-level. The L1 SPM is private to each
cluster and split into twice as many banks as cores to keep contention low
(typically $<$10\%). Cores can read and write to L1 SPM within a single cycle
and use it as low-latency shared memory and for fine-grained synchronization.
Each cluster contains a multi-channel DMA for efficient data movement.
%

\subsection{Throughput Considerations}
The discussed configuration runs at up to 1\,GHz in the target technology, and
hence the raw compute throughput amounts to 32\,Gop/s, fulfilling the
requirement of 2\,Gop/s and providing ample headroom for cases where more
compute power is needed instantaneously (e.g., bursts of complex datatypes).
Further, the memories and system interconnects have been sized to be 256\,bit
wide to sustain a line rate of 200\,Gbit/s.

New packets enter the accelerator over a 256\,bit wide input port on the
network side, and are first placed in the L2 SPM. The cluster-private DMAs then
transfer the data to the local L1 SPMs. After processing, the results are
directly transferred from the L1 SPMs to the 256\,bit output port on the PCIe
side. Since the L2 needs to sustain a bidirectional bandwidth of
2$\cdot$256\,Gbit, it has been split into two banks with separate ports into
the system interconnect. Each cluster processes, on average, one fourth of the
incoming data, and hence each DMA can transfer 64\,bit/cycle in each direction.
%

\subsection{Cycle Accurate Simulations}%
The cycle-accurate testbed is comprised of synthesizable SystemVerilog models
that are simulated with Mentor QuestaSim. The NIC/PCIe interfaces are abstracted
as a bus-attached memory region and a FIFO, respectively. 
%
\vspace{-0.2em}\subsubsection{Bandwidth} To measure the effective bandwidth of memories,
interconnect, and DMA engines, we created a benchmark where cores continuously
transfer data blocks from L2 to the local L1 and then to the host PCIe using
DMA bursts.  The block size is varied from 256\,B to 128\,KiB.  As shown in
Fig.~\ref{fig:hw}c), a throughput of 192\,Gbit/s can be reached for blocks of
256\,B, and all higher block sizes are above the line rate.

\vspace{-0.2em}\subsubsection{Microkernels}\label{sec:microkernels}
To study the performance of
datatype processing on PULP, we benchmark the RW-CP handlers on a 
$1$ MiB message with a vector datatype, varying the block size.
Fig.~\ref{fig:vec-hw} shows the throughput achieved by simulating DDT 
processing on PULP (cycle-accurate) and compares it with the ones of  
a ARM-based simulation (SST+gem5, see Sec.~\ref{sec:sim-testbench}).

The benchmark creates dummy packets and HERs in the L2 memory and statically
assigns the HERs to the cores. To emulate the blocked-RR scheduling strategy
used by RW-CP, we assign consecutive blocks of $4$ packets ($2$ KiB/packet) to
each core. We report the achieved throughput based on the maximum time needed
by a core to process all the assigned packets. We keep the dataloops (i.e.,
the datatype description) in L2 and move the checkpoints into the L1 of 
the cluster where the handlers using them are run.
\begin{figure}[h]
    \vspace{-0.5em}
    \includegraphics[width=1\columnwidth]{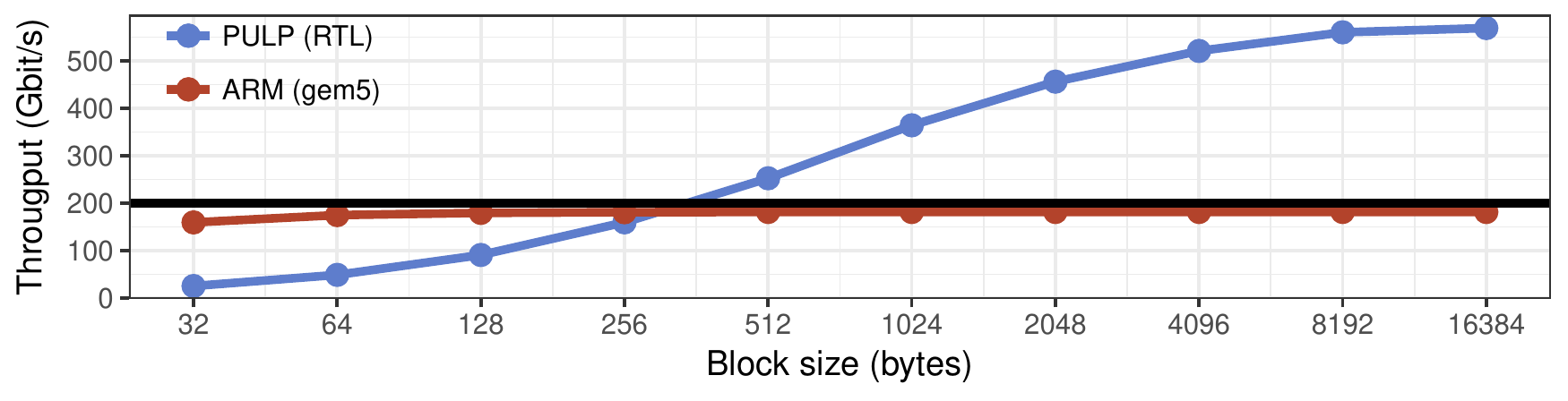}
    \vspace{-2.5em}
    \caption{DDT processing throughput on ARM and PULP.}
    \vspace{-1em}
    \label{fig:vec-hw}
\end{figure}

The PULP-based implementation is slower than the ARM-based one for small block
sizes (i.e., < $256$ B): in these cases the handlers run longer because of the
higher number of contiguous blocks per packet, making more accesses to L2, thus
increasing memory contention. However, we have also to take into account that
the gem5-based simulation may not model memory contention properly, or at least
not as good as the cycle accurate simulation we use for PULP.  The impact of L2
contention is visible also in Fig.~\ref{fig:vec-ipc-hw}, that shows the
instructions-per-cycle (IPC) of the RW-CP handlers on PULP: small block sizes
lead to lower IPCs due to the more frequent L2 accesses.  The PULP-based DDT
processing achieves line rate for block sizes larger than $256$ bytes. After
that, line rate is exceeded because the experiment is not capped by the network
bandwidth (the packets are preloaded in L2).  We reserve a more in-depth
evaluation and optimization (e.g., transparently moving dataloops to L1) of a
sPIN implementation on PULP for future work.

\begin{figure}[h]
    \vspace{-0.5em}
    \includegraphics[width=1\columnwidth]{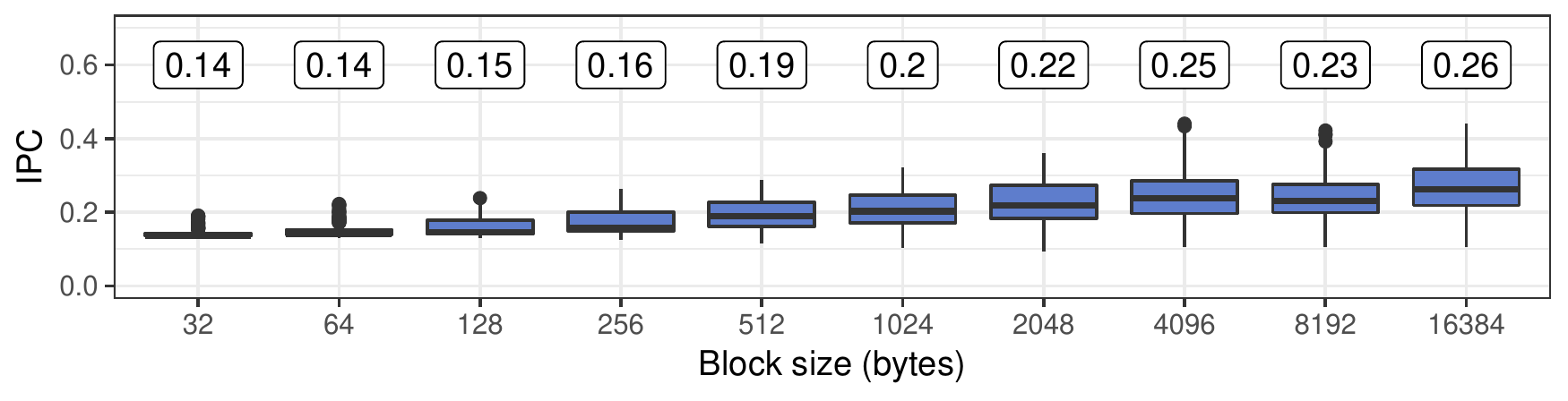}
    \vspace{-2.5em}
    \caption{RW-CP Instructions per Cycle (IPC) on PULP. The median IPC for
each block size is annotated on the top.}
    \vspace{-1em}
    \label{fig:vec-ipc-hw}
\end{figure}

%
%
%

\vspace{-0.5em}
\subsection{Circuit Complexity and Power Estimations}%
\label{sec:proc:22fdsoi}
We synthesized the processor in GlobalFoundries' 22\,nm fully depleted silicon
on insulator (FDSOI) technology using Synopsys DesignCompiler, and were able to
close the timing of the system at 1\,GHz. Including memories, the entire
accelerator has a complexity of approximately 100\,MGE \footnote{1 gate
equivalent (GE) equals 0.199\,$\mu$m$^2$, the area of two-input NAND gate in
22\,nm.}.  Fig.~\ref{fig:hw}b) shows the area breakdown of the entire
accelerator (left) and of one processing cluster (right). The four clusters
together account for ca.\ 39\% of the total accelerator area, and the remaining
part is mainly dominated by the 8\,MiB of L2 SPM (59\%). The system
interconnect, DWCs, and buffers only amount to around 2\%.  Within one cluster,
the 1\,MiB L1 SPM accounts for 84\% of the area, the rest is taken by the
shared instruction cache (7\%), the eight RISC-V cores (6\%), and the DMA
engine and interconnects (3\%). Assuming a conservative layout density of 85\%,
100\,MGE equal 23.5\,mm$\sp{2}$ of silicon area.

To put this in perspective, we compare the silicon area of the proposed \spin
accelerator with the compute subsystem of the BlueField SmartNIC system-on-chip
(SoC) from Mellanox~\cite{blueField2019}. These devices contain two 100\,Gbit
Ethernet ports, a Gen4 16$\times$ PCIe port and an additional ARM subsystem
with up to 16 A72 64\,bit cores with 1\,MiB shared L2 cache per 2 cores, and
6\,MiB shared L3 last-level cache. From \cite{Mair2016,Pyo2015} it can be
inferred that one A72 dual-core tile occupies around 5.6\,mm$^2$ in 22\,nm
technology, and hence a full 16 core configuration amounts to around
51\,mm$^2$. Hence, with an overall complexity of 23.5\,mm$\sp{2}$, we observe
that the analyzed parameterization of the proposed accelerator only occupies
about 45\% of the area budget allocated for the processing subsystem in this
related commercial device. Note that the presented architecture is modular and
could be re-parameterized based on the technology node, available silicon area
and application needs. E.g., with a similar area budget as on the BlueField
SoC, we could double the amount of clusters and memory to 64 RISC-V cores and
18\,MiB.

\sloppypar{
In terms of power consumption, we estimate (using a back-annotated gate-level
power simulation) that this design requires around 6\,W under
full load (not including I/O and PHY power), which is comparable to the typical power consumption that has been measured for PCIe NIC cards (4.6-21.2\,W)~\cite{Sohan2010}.}

\vspace{-0.5em}
\subsection{Discussion}
The PULP platform well matches the sPIN abstract machine, making it a
first-class candidate for a sPIN implementation.  The main research directions
we plan to explore in this context are: (1) Extend the sPIN programming model in
order to let the user specify which data should be moved to L1. In fact, as
discussed in Sec.~\ref{sec:microkernels}, L2 contention can quickly become an
issue for performance. (2) Design a sPIN runtime running on PULP. The runtime
is in charge to manage the cores/clusters, assingning new HERs to execute to the idle
ones, and serve the commands issued by the handlers (e.g., DMA read/writes, new
network operations). (3) Deeply evaluate further use cases in order to validate
our design choices.




\vspace{-0.5em}
\section{Evaluation}
\label{sim}
%
We evaluate the effects of datatypes processing offload using an extensive set
of simulations on a modified version of Cray Slingshot Simulator.
We now describe the simulation setup, then we discuss the
microbenchmarks that are used to study the performance and NIC resource usage
for DDT offloading solutions. Finally, we show the performance improvements
that can be achieved on full HPC applications.  Hardware implementation aspects
of \spin will be discussed in Sec.~\ref{sec:hw}.

\subsection{Simulation Setup} \label{sec:sim-testbench}

To analyze the effects of integrating \spin in next-generation networks, we
extend the Cray Slingshot Simulator, which models a $200$ Gib/s NIC (see
Fig.~\ref{fig:spin-model}) in Sandia Structural Simulation Toolkit
(SST)~\cite{janssen2010simulator}, adding packet processing capabilities by
implementing \spin. We configure the network simulator to send $2KiB$ of payload
data.
We combine the network simulator with gem5~\cite{binkert2011gem5}, that
accurately simulates ARM-based architectures~\cite{endo2014micro}, to simulate
the handlers execution.
The HPUs are modeled as 64-bit A15 processors~\cite{tousi2017arm}.  We associate
each NIC with a gem5 system, configured with $32$ Cortex A15 clocked at
$800$ MHz (unless otherwise specified). The NIC memory is simulated with the
gem5's SimpleMemory module, that can support k-cycles latency memory accesses
(we use $k=1$ in this paper) with a bandwidth of $50$ GiB/s and a number of
channels equal to two times the number of HPUs.
The host-NIC interface is modeled as a x32 PCIe Gen4: the simulation accounts
for PCIe packets overheads and a 128b/130b encoding scheme.  The SST-gem5
integration follows the same approach adopted by Hoefler et al. to simulate
\spin handlers in LogGOPSim~\cite{hoefler-loggopsim}.

The host-based unpack function is profiled by running the
MPITypes unpack function (i.e., \texttt{MPIT\_Type\_memcpy}) on a Intel i7-4770
CPU @3.4GHz equipped with 32\,GiB of DRAM. Each measure is taken until the
$99$\% CI is within the $10$\% of the reported median. For RW-CP we set $\epsilon = 0.2$.  (see
Sec.~\ref{sec:general_handlers}), so to keep the scheduling overhead less than
20 \% of the message processing time.

\subsection{Microbenchmarks}

The microbenchmarks presented in this section are based on the offloading
of an MPI vector datatype. We observe similar results for other MPI derived
datatypes but omit them due to space limitations.

\begin{figure}[h]
    \vspace{-0.5em}
    \centering{\includegraphics[trim=0 0 0 0, clip,
    width=1\columnwidth]{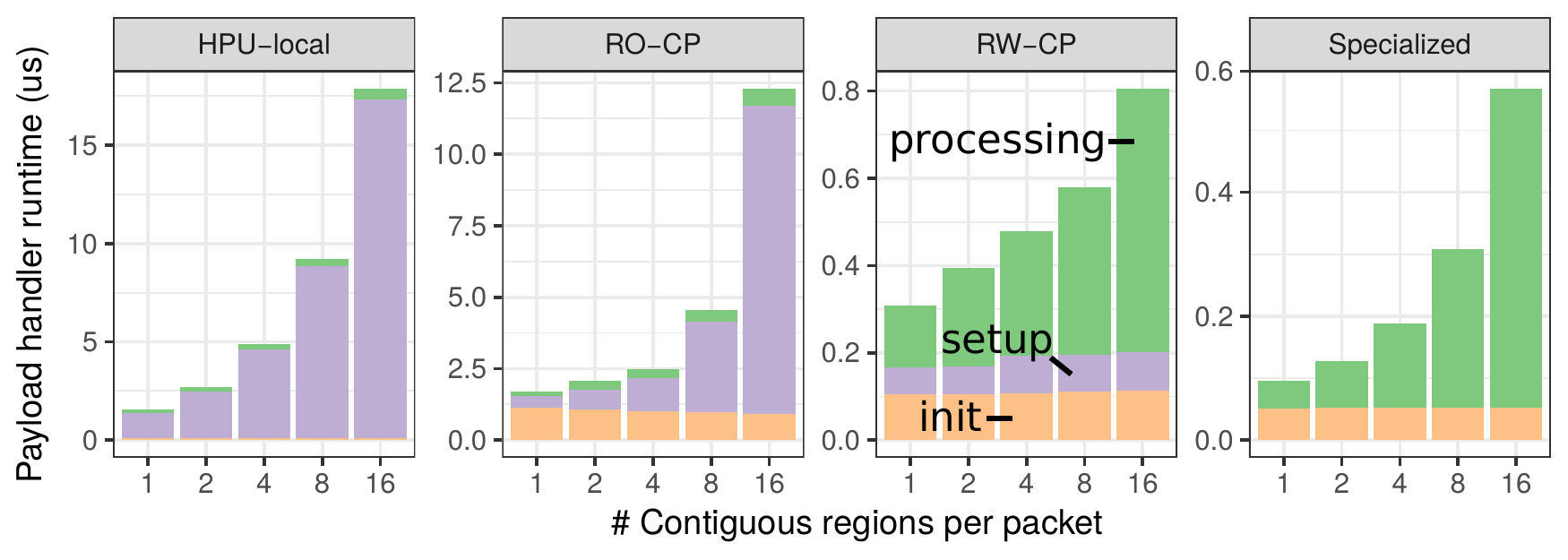}}
    \vspace{-2.3em}
    \caption{Payload handlers execution breakdown for different strategies
    and block sizes.}
    \vspace{-1em}
    \label{fig:microbench_vec_handlers}
\end{figure}

\paragraph{Unpack Throughput.} Fig.~\ref{fig:microbench_vec_block} shows the
receive throughput for a 4\,MiB message described with a vector datatype,
varying the block sizes. This value is computed considering the time from when
the \textit{ready-to-receive} message is sent to the sender (i.e., to avoid
that data message arrives unexpectedly) to when the last byte of data is copied
into its final position in the receiver memory.

For small block sizes, the offloaded approaches lose their
effectiveness, becoming more expensive than the host-based unpack for $4$ byte
blocks. This behavior is explained by Fig.~\ref{fig:microbench_vec_handlers},
that shows a breakdown of the handlers runtime for small block sizes.  We
report the handlers performance as function of $\gamma$, i.e., the number of
contiguous regions per packet, varying it from $1$ (i.e., block size of $2048$\,B)
to $16$ (i.e., block size of $128$ B).

The HPU-local handler runtime is always dominated by the \textit{setup} time
(including the MPITypes catch-up phase). RO-CP spends more time in the
\textit{init} phase, where the checkpoint gets copied, and is also involved in
long catch-up phases (e.g., 87\% of the total time for $\gamma=16$).
RW-CP is only a factor of two slower than the specialized handler, explaining
the high throughput reached in Fig.~\ref{fig:microbench_vec_block}.

\paragraph{Scalability.}
Fig.~\ref{fig:microbench_vec_mix}a shows the receive throughput for different
numbers of HPUs, fixing the block size to $2$ KiB (i.e., $\gamma=1$). The
specialized handler reaches line rate already with two HPUs, while the others
are limited by the respective overheads (i.e., scheduling for RW-CP, catch-up
for HPU-local, copy and catch-up for RO-CP).

\begin{figure}[t]
    \centering{\includegraphics[trim=0 0 0 0, clip,
    width=1\columnwidth]{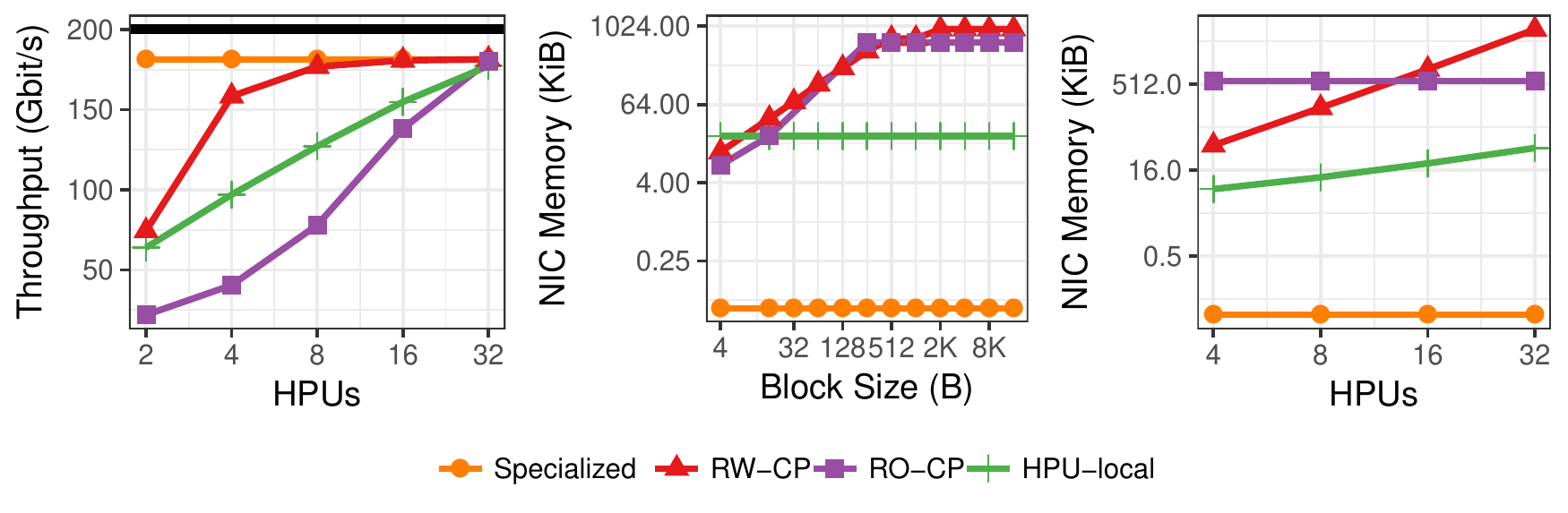}}
    \vspace{-2.5em}
    \caption{From left to right: Receive throughput scalability; NIC Memory 
            occupancy as function of the block size; NIC Memory occupancy as 
            function of the number of HPUs.}
    \vspace{-1em}
    \label{fig:microbench_vec_mix}
\end{figure}
Fig.~\ref{fig:microbench_vec_mix}b shows the NIC memory occupancy for different
block sizes, fixing the number of HPUs to $16$. While the HPU-local and
specialized occupancy remains fixed, the checkpointed variants adjust the
checkpoint interval to keep their scheduling overhead less than $\epsilon$, as
described in Sec.~\ref{sec:general_handlers}. In particular, the larger the
block size, the faster the message processing time, the smaller will be the
checkpoint interval (leading to higher NIC memory occupancy).

Fig.~\ref{fig:microbench_vec_mix}c shows the NIC memory occupancy for
different number of HPUs. The memory resources required by the HPU-local 
strategy increase
with the number of HPUs because the segment data structure is replicated
on each HPU. For RW-CP, by increasing the
number of HPUs we make the DDT processing faster, leading the
checkpoint interval selection heuristic to increase the number of checkpoints,
thus the higher NIC memory occupancy.

\begin{figure}[h]
    \vspace{-0.8em}
    \centering{\includegraphics[trim=0 0 0 0, clip,
    width=1\columnwidth]{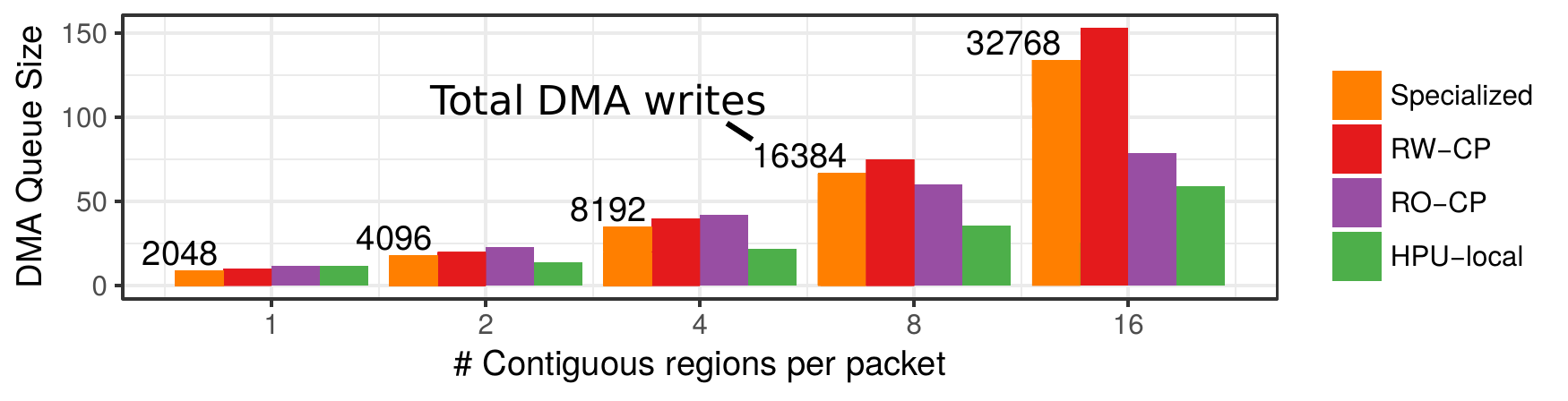}}
    \vspace{-2.5em}
    \caption{Maximum DMA write requests queue occupancy over the entire
    message processing time.}
    \vspace{-1em}
    \label{fig:microbench_vec_dma_queue}
\end{figure}

\paragraph{PCIe Traffic.}
In this microbenchmark we study the PCIe traffic generated by different
handlers.  Fig.~\ref{fig:microbench_vec_dma_queue} shows the maximum DMA queue
occupancy that has been reached during the message processing, for different
values of $\gamma$ (x-axis) and different handlers (bars). Each group of bars
is annotated with the total number of DMA writes that are issued for that
specific value of $\gamma$. The number of HPUs is fixed to $16$.
In all the cases, the PCIe request buffer is kept under 160 requests, meaning
that PCIe was not a bottleneck.

\begin{figure}[h]
    \vspace{-0.8em}
    \centering{\includegraphics[trim=0 0 0 0, clip,
    width=1\columnwidth]{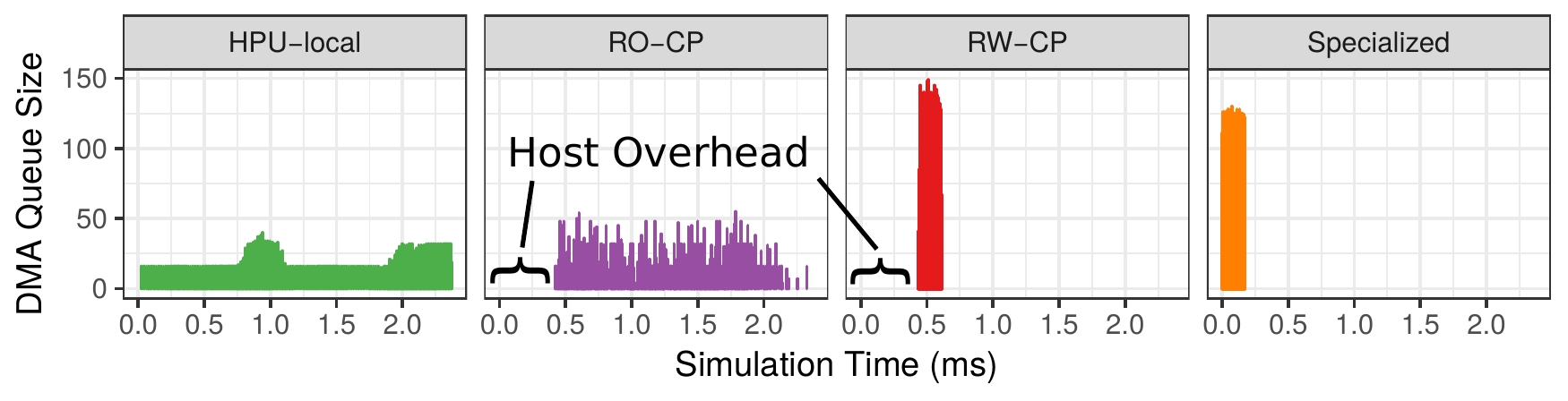}}
    \vspace{-2.8em}
    \caption{DMA write requests queue over time for $\gamma=16$ and different
    datatype processing strategies.}
    \vspace{-1em}
    \label{fig:microbench_vec_dma_queue_detail}
\end{figure}
Fig.~\ref{fig:microbench_vec_dma_queue_detail} shows the DMA FIFO queue size
over time for handlers processing a message with $\gamma=16$ (i.e., $128$ B
block sizes). The HPU-local and RO-CP strategy have the slowest handlers,
which translates to a small number of DMA requests issued per second,
explaining a low DMA queue occupancy. 
The RW-CP and specialized strategies have higher peak occupancy
due to their faster handlers. The figure also shows the time needed by
the host to create the checkpoints and copy them to NIC memory (i.e.,
\textit{host overhead}).
%

\vspace{-0.5em}
\subsection{Real Applications DDTs}


\begin{figure*}[h!]
    \includegraphics[width=1\textwidth, trim={0 15 0 0}, clip]{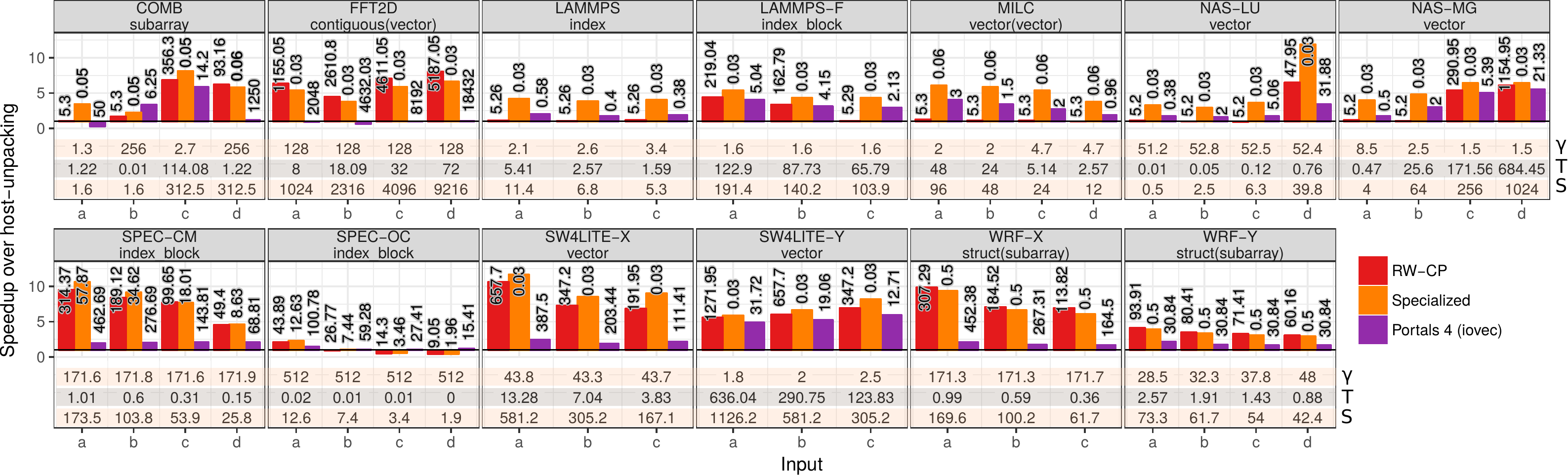}
    \vspace{-0.7cm}
    \caption{Message processing time speedup for
different DDT kernels. For each input, we report \textbf{$\gamma$}:
average number of blocks per packet; \textbf{T}: the baseline, that is the message processing time
for the host-based unpacking (ms); \textbf{S}: the message size (KiB). The bars are annotated
with the amount of data (KiB) moved to the NIC to support the unpack. }
\vspace{-1em}
\label{fig:microapps}
\end{figure*}

We analyze a set of datatypes used in real applications and for different input
parameters. The selected applications cover different domains such as
atmospheric sciences, quantum chromodynamics, molecular dynamics, material
science, geophysical science, and fluid
dynamics~\cite{schneider-app-oriented-ping-pong}.
The selected applications are:
\begin{itemize}[noitemsep,topsep=0pt,parsep=0pt,partopsep=0pt,leftmargin=*]

\item \textbf{COMB}~\cite{comb}: a communication performance benchmarking
tool for common communication patterns on HPC platforms. The tool represents
exchange of data stored in n-dimensional arrays that are distributed across one
or more dimensions.

\item \textbf{FFT2D}~\cite{hoefler-datatypes}: performs a Fast
Fourier Transform (FFT) for a 2D distributed matrix by row-column algorithm which
composes 2D transform from two one-dimensional FFT applied to rows and then to
columns. Such algorithm requires transposing the matrix which can be expressed
as a non-contiguous data access.

\item \textbf{LAMMPS}~\cite{Plimpton:1995:FPA:201627.201628}: a molecular
dynamics simulator for materials modeling.  The application
exchanges the physical properties of moving  particles using index datatypes.
Depending on simulation settings, the particles can have different number of
properties, therefore, we create two tests: \textit{LAMMPS\_full} and
\textit{LAMMPS}.

\item \textbf{MILC}~\cite{bernard1991studying}: a large scale numerical
simulation for quantum chromodynamics.  We use the lattice
quantum chromodynamics which performs halo-exchange of 4D regular
grid of points.

\item \textbf{NAS}~\cite{van2002parallel}: a fluid dynamics simulation system
solving Navier-Stokes equations. It consists of two micro-applications: NAS\_MG
and NAS\_LU. NAS\_MG communicates the faces of a 3D array, and
NAS\_ LU solves a three-dimensional system of equations.

\item \textbf{SPECFEM3D}~\cite{Carrington:2008:HSG:1413370.1413432}: simulates
seismic wave propagation problems using Finite
Element Method.  It uses indexed datatypes for exchanging mesh
grid points with neighboring processes and exhibits two different exchange
patters (\textit{FEM3D\_\{oc,cm\}}), which differ in the amount of data
communicated per index.

\item \textbf{SW4LITE}~\cite{sw4lite}: a performance testing library which
solves seismic wave equations in Cartesian coordinates for 3-D seismic
modeling. It uses different datatypes for exchanging data with
neighboring processes: we distinguish between exchanges along x and
y directions, and define two tests: \textit{sw4\_x} and \textit{sw4\_y}.

\item \textbf{WRF}~\cite{Skamarock:2008:TNA:1347465.1347775}: a numerical
weather prediction system. It represents the space as 3 dimensional Cartesian
grid, and performs halo exchanges of structs of subarray datatypes. We define
two tests depending on the exchange direction:  \textit{wrf\_x} and
\textit{wrf\_y}.
\end{itemize}

Figure~\ref{fig:microapps} shows the speedup of the RW-CP and specialized handlers
w.r.t. the host-based unpack for different datatypes and
message sizes employed by the above described applications.
The host-based unpacking receives the full message and then unpacks it with
MPITypes. The benchmark is executed with cold caches to model the scenario
where the message has just been copied from the NIC to main memory (we assume
no direct cache placement of the DMA writes).
We compare the achieved speedups with a Portals 4-based unpacking solution.
This solution uses input/output vectors (iovecs) and assumes the NIC being able
to store a number $v$ of iovec entries: Every received $v$ blocks, the NIC
issues a PCIe read (modeled with a $500ns$ latency~\cite{neugebauer2018understanding, pciecong}) from main memory to get the
next $v$ iovecs. We use $v=32$, that is the maximum number of scatter-gather
entries for a Mellanox ConnectX-3 card~\cite{core-direct} (no Portals 4
implementation is publicly available at the time of writing). This model
assumes in-order packet arrival.

For each experiment, we report the average number of blocks per packet
($\gamma$), the message processing time of the host-based solution (T), and the
message size (S). The bars are annotated with the size of data moved to the NIC
in order to offload DDT processing: RW-CP needs to copy the MPITypes
dataloops describing the datatype and the checkpoints (see
Sec.~\ref{sec:general_handlers}); the specialized handler always require the
minimum amount of space (e.g., the list of offsets and block sizes for the
index datatype); the Portals 4 solution needs to move the entire iovec list,
which size is linear in the number of contiguous regions identified by the
datatype.

RW-CP and native can reach up to 12x speedup over the host-based unpacking.
The offloaded datatype processing strategies do not introduce any speedup in
the cases where the message size is small (i.e., the first two COMB experiments
send messages fitting in one packet) or if the number of blocks per packet is
large (e.g., SPEC-OC has $\gamma=512$ blocks per packet). 
In the first case, the datatype processing cannot exploit the NIC parallelism
and the datatype processing is only delayed by the handler latency. 
In the second case, the message processing time increases because of: (1) the
increased handler runtime, that is linear in the number of contiguous regions ;
(2) the inefficient utilization of the PCIe bus (i.e., with $\gamma=512$, the
handlers issue $512$ DMA writes of $4$ bytes). 
%

\paragraph{Memory Transfers.}
Host-based message unpacking requires the NIC to first write the received
(packed) message in a memory buffer so the CPU can unpack it. During the
unpack, the CPU needs to access the entire packed data and all the parts of the
receive buffer where the message has to be copied to. Instead, by offloading
the datatype processing task, the only memory accesses that are needed are the
ones that the NIC does to write the data directly in the receive buffer. In
Fig.~\ref{fig:mtransfers} we show the total data volume that is moved to and
from the main memory to receive and unpack a message for RW-CP and host-based unpacking. 

\begin{figure}[h]
    \includegraphics[width=1\columnwidth]{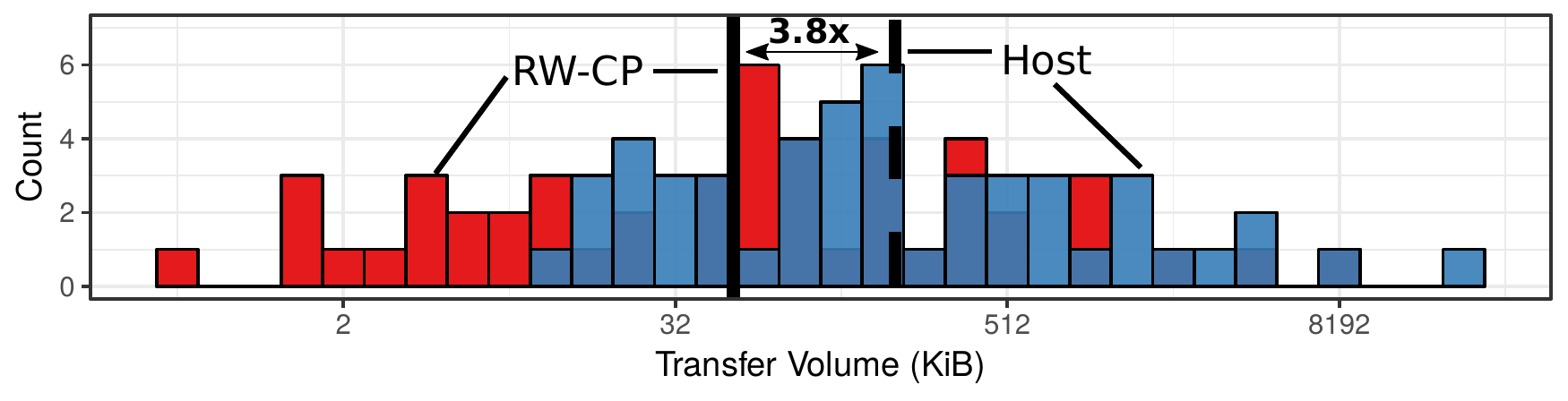}
    \vspace{-2.3em}
    \caption{Histogram of the data volumes moved to/from main memory by
    RW-CP and host-based unpacking for the experiments of Fig.~\ref{fig:microapps}.
    The vertical lines represent the geometric mean of the moved data volumes for
    RW-CP (continuous line) and host-based unpacking (dashed line).}
\vspace{-1.3em}
\label{fig:mtransfers}
\end{figure}

We report how many experiments of Fig.~\ref{fig:microapps} (y-axis) needed to move a
specific amount of data (x-axis) for completing the message unpacking. For
RW-CP, the data volume written to main memory always corresponds to the message
size. For the host-based unpack, the data volume is the message size (moved
from the NIC to the main memory) plus the data transferred between the
last-level cache and the main memory during the unpack (measured as
number of last-level cache misses times the cache line size).  
The reported geometric mean shows that RW-CP moves $3.8x$ less data than the host-based unpack.

\paragraph{Amortizing Checkpointing Cost.}
The RW-CP strategy requires the application to create checkpoints before starting
to receive the data to unpack. But, \textit{how long does it take to produce and
copy these checkpoints to the NIC?} 
We answer this question with the data shown in Fig.~\ref{fig:checkpoint_cost}.
Instead of reporting an absolute number, we report for each application/input
combination of Fig.~\ref{fig:microapps} the number of times a datatype should
be used in order to amortize the checkpoint creating cost.
\begin{figure}[b]
    \vspace{-1em}
    \includegraphics[width=1\columnwidth]{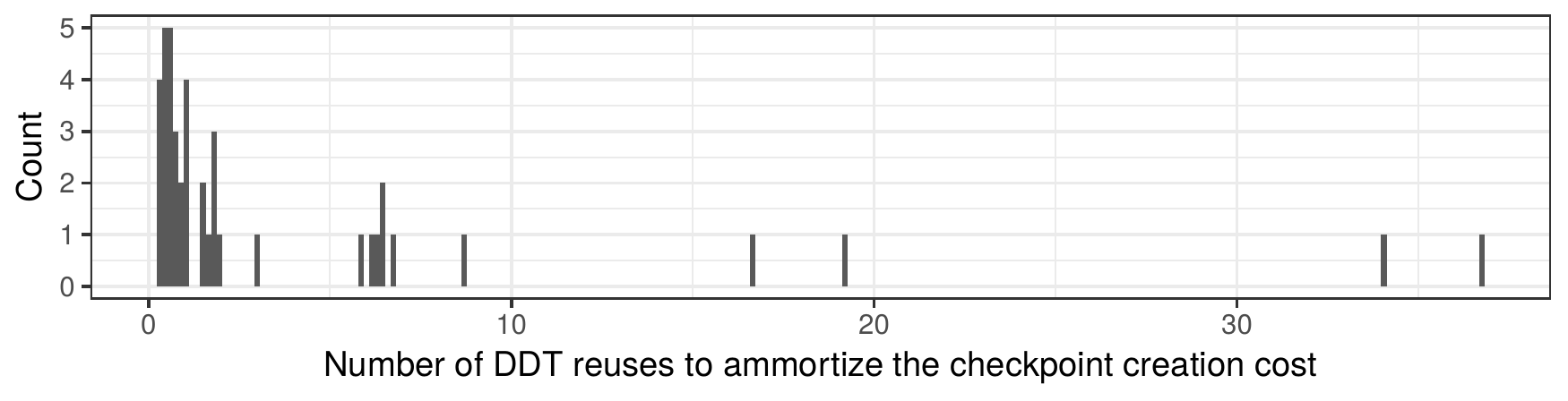}
    \vspace{-2.3em}
    \caption{Number of times a datatype needs to be reused to amortize the checkpoint
creation overhead for RW-CP.}
\label{fig:checkpoint_cost}
\end{figure}
In the $75\%$ of the cases, the speedup introduced by RW-CP pays off after $<4$
reuses of the same datatype.
It is worth noting that the checkpoints are independent from the receive buffer
(i.e., they are used to compute offsets), hence the checkpoint creation cost is
paid only once per datatype. With iovecs, instead, each iovec entry needs to
have the virtual address where the block starts, hence a iovec list needs to be
created every time the receive buffer changes.

\subsection{Application Scalability}
To study the effects of offloading datatype processing we benchmark the full
FFT2D application at large scales. The application partitions the input
matrix by rows and performs two 1D-FFT operations. The second one is applied after
the matrix is transposed with a \texttt{MPI\_Alltoall} operation. After the
second 1D-FFT finishes, the matrix is transposed back to the original layout.
We use the same approach of Hoefler et al.~\cite{hoefler-datatypes}, that avoids the manual
matrix transposition by encoding this operation as MPI Datatypes.

We simulate the effect of offloading the datatype processing (hence the matrix 
transposition) to sPIN at large scales. In particular, given a communicator size,
we simulate the unpack cost of the receive datatype with the SST and measure 
the 1D-FFT time for the different workloads (see Sec.~\ref{sec:sim-testbench} for
the detailed configuration). We use these two parameters to build a GOAL~\cite{hoefler-goal}
trace for FFT2D at different scales. We then run the trace with LogGOPSim, that is
configured to model next-generation networks~\cite{spin}.

\begin{figure}[h]
    \vspace{-1em}
    \includegraphics[width=1\columnwidth]{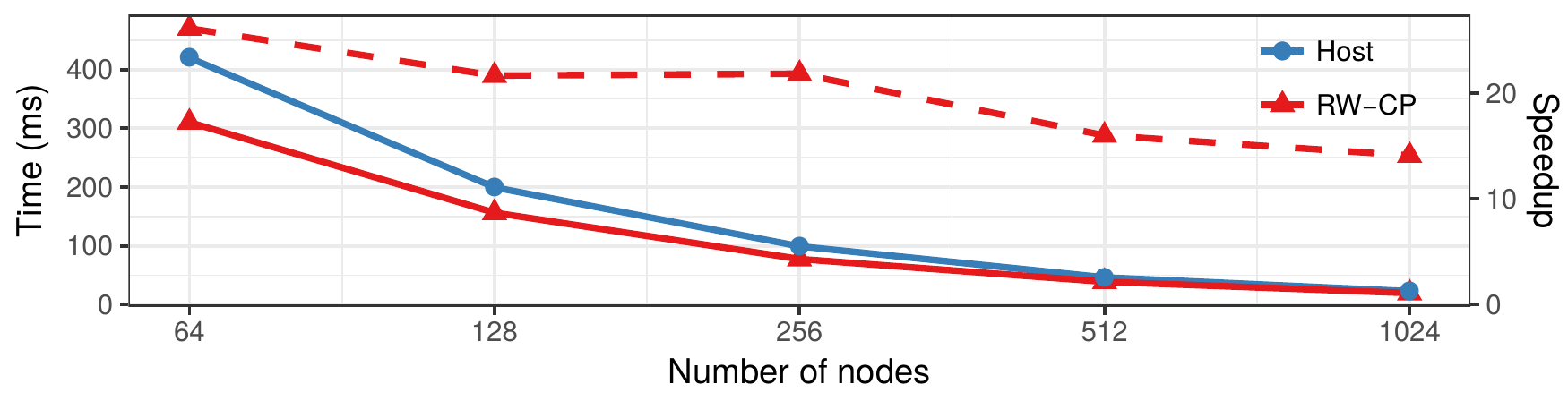}
    \vspace{-2.3em}
    \caption{FFT2D strong scaling: runtime (continuous) and
speedup (\%, dashed) of RW-CP over host-based unpacking.}
\vspace{-1em}
\label{fig:fft_scal}
\end{figure}
Fig.~\ref{fig:fft_scal} shows the FFT2D strong scaling for a matrix $n \cdot n$
with $n=20,480$. At $P=64$ the runtime of each node is split in $\sim60\%$
computation and $\sim40\%$ communication: offloading datatype processing
accelerates the application up to $26\%$ of the host-based unpack version (i.e.,
MPITypes). Increasing the number of nodes, the unpack overhead shrinks, 
reducing the effects of optimizing it.



%
\section{Related Work}
We discussed the various solutions to support non-contiguous memory
transfers in Sec.~\ref{sec:ncmt}. Several works have been focusing on
optimizing MPI datatypes with different approaches: Traff et
al.~\cite{10.1007/3-540-48158-3_14} show that some complex derived datatypes
can be transformed into simpler ones, improving the packing and unpacking
performance with a more compact representation.  Gropp et
al.~\cite{gropp1999improving} provide a classification of the MPI derived
datatypes based on their memory access patterns, discussing how they can be
efficiently implemented or automatically optimized~\cite{byna2006automatic}.
Both approaches are orthogonal to this work because they propose
optimizations that can be applied before offloading and can be integrated
in the offloaded handlers.

Offloading DDT processing to GPU systems has been discussed by Wang et
al.~\cite{panda2}. However, unpacking data on GPUs still requires the data
to be fully received on the GPU memory before the unpack can start. 
In this work, we discuss how datatype
processing can be performed as the stream of data arrives, truly implementing
zero-copy non-contiguous memory transfers.

Non-contiguous memory transfers can be accelerated by compiling DDTs
pack/unpack functions at runtime~\cite{schneider-rtcompmpiddt,prabhu2015dame}.
However, the main optimization of this approach it to choose the best data
copying strategy for x86 architectures~\cite{schneider2013compiler}. The same
cannot be applied to NICs, where the only way to move data to the host is by
issuing DMA writes to it.  The performance we expect from runtime-compilation
is the same of the specialized handlers shown in this work.  Our
interpretation-based offload approach (i.e., RW-CP) has lower startup cost,
since there is no compilation step necessary, and shows performances similar to
the specialized handlers.

The idea of triggering handlers in response to incoming data has been explored
with Active Messages~\cite{eicken1992active, besta-aam} for message-passing and
with Active Accesses~\cite{besta-aa} for one-sided operations. While these
solutions can be used to trigger datatype unpacking routines, they only express
handlers as function of the received message (i.e., no streaming processing)
and they are run on the host CPU. In sPIN the handlers are defined on a
per-packet basis and run directly on the NIC, enabling datatype processing on
the fly.  Di Girolamo et al.~\cite{salvooff} propose an abstract machine model
to express offloaded communication/computation schedules: DDT processing can be
modeled with such schedules (i.e., triggered puts targeting the local node to
scatter the data). However, also in this case the processing can only start
after the full message has been received and it requires a number of NIC resources
linear in the number of contiguous blocks. 


\section{Conclusions}

We presented different solutions to accelerate non-contiguous memory transfers
exploiting next-generation network accelerators like \spin. We identify what are
the challenges of offloading MPI derived datatypes, the most
expressive interface to describe non-contiguous memory regions, and how \spin can be
extended to efficiently support on-the-fly datatype processing.  To back our
simulations studies, we outline a real hardware implementation of \spin. We
then provide an extensive evaluation showing how communications can benefit from
network-accelerated datatype processing.

\subsection*{Acknowledgements}
This project has received funding from the European Research Council (ERC)
under the European Union's Horizon 2020 programme (grant agreements: DAPP, No.
678880; OPRECOMP, No. 732631).
Jakub Ber\'anek has been supported by the European Science Foundation through
the ``Science without borders'' project, reg. nr.
CZ.02.2.69/0.0./0.0./16\_027/0008463 within the Operational Programme Research,
Development and Education.
We thank David Hewson for the insightful discussions and Cray Inc. for
supporting this project.

\balance
{
\setlength{\parskip}{0pt}
\setlength{\bibsep}{0pt plus 1ex}
\bibliographystyle{ACM-Reference-Format-abbrv}
\bibliography{bibliography}}


\begin{thebibliography}{00}


\ifx \showCODEN    \undefined \def \showCODEN     #1{\unskip}     \fi
\ifx \showDOI      \undefined \def \showDOI       #1{{\tt DOI:}\penalty0{#1}\ }
  \fi
\ifx \showISBNx    \undefined \def \showISBNx     #1{\unskip}     \fi
\ifx \showISBNxiii \undefined \def \showISBNxiii  #1{\unskip}     \fi
\ifx \showISSN     \undefined \def \showISSN      #1{\unskip}     \fi
\ifx \showLCCN     \undefined \def \showLCCN      #1{\unskip}     \fi
\ifx \shownote     \undefined \def \shownote      #1{#1}          \fi
\ifx \showarticletitle \undefined \def \showarticletitle #1{#1}   \fi
\ifx \showURL      \undefined \def \showURL       #1{#1}          \fi
\providecommand\bibfield[2]{#2}
\providecommand\bibinfo[2]{#2}
\providecommand\natexlab[1]{#1}
\providecommand\showeprint[2][]{arXiv:#2}

\bibitem[\protect\citeauthoryear{Shanley}{Shanley}{2003}]%
        {IB}
\bibfield{author}{\bibinfo{person}{T. Shanley}.}
  \bibinfo{year}{2003}\natexlab{}.
\newblock \bibinfo{booktitle}{{\em Infiniband Network Architecture}}.
\newblock \bibinfo{publisher}{Addison-Wesley Professional}.
\newblock


\bibitem[\protect\citeauthoryear{??}{cor}{2019}]%
        {core-direct}
 \bibinfo{year}{2019}\natexlab{}.
\newblock \bibinfo{title}{{Mellanox Technologies}}.
\newblock \bibinfo{howpublished}{\url{http://http://www.mellanox.com/}}.
  (\bibinfo{year}{2019}).
\newblock


\bibitem[\protect\citeauthoryear{Alverson, Froese, Kaplan, and Roweth}{Alverson
  et~al\mbox{.}}{2012}]%
        {Aries}
\bibfield{author}{\bibinfo{person}{B. Alverson}, {et~al\mbox{.}}}
  \bibinfo{year}{2012}\natexlab{}.
\newblock \showarticletitle{Cray XC series network}.
\newblock \bibinfo{journal}{{\em Cray Inc., White Paper WP-Aries01-1112\/}}
  (\bibinfo{year}{2012}).
\newblock


\bibitem[\protect\citeauthoryear{Barrett, Brightwell, Hemmert, Pedretti,
  Wheeler, Underwood, Riesen, Hoefler, Maccabe, and Hudson}{Barrett
  et~al\mbox{.}}{2018}]%
        {portals42}
\bibfield{author}{\bibinfo{person}{B.~W Barrett}, {et~al\mbox{.}}}
  \bibinfo{year}{2018}\natexlab{}.
\newblock \showarticletitle{{The Portals 4.2 Network Programming Interface}}.
\newblock \bibinfo{journal}{{\em Sandia National Laboratories, November 2012,
  Technical Report SAND2012-10087\/}} (\bibinfo{year}{2018}).
\newblock


\bibitem[\protect\citeauthoryear{Schneider, Hoefler, Grant, Barrett, and
  Brightwell}{Schneider et~al\mbox{.}}{2013}]%
        {schneider-portalsoffload}
\bibfield{author}{\bibinfo{person}{T. Schneider}, {et~al\mbox{.}}}
  \bibinfo{year}{2013}\natexlab{}.
\newblock \showarticletitle{{Protocols for Fully Offloaded Collective
  Operations on Accelerated Network Adapters}}. In \bibinfo{booktitle}{{\em
  Parallel Processing (ICPP), 2013 42nd International Conference on}}.
  \bibinfo{pages}{593--602}.
\newblock
\showISSN{0190-3918}


\bibitem[\protect\citeauthoryear{Girolamo, Jolivet, Underwood, and
  Hoefler}{Girolamo et~al\mbox{.}}{2016}]%
        {salvooff}
\bibfield{author}{\bibinfo{person}{S.~Di Girolamo}, {et~al\mbox{.}}}
  \bibinfo{year}{2016}\natexlab{}.
\newblock \showarticletitle{{Exploiting Offload Enabled Network Interfaces}}.
\newblock \bibinfo{journal}{{\em IEEE MICRO\/}} \bibinfo{volume}{36},
  \bibinfo{number}{4} (\bibinfo{date}{Jul.} \bibinfo{year}{2016}).
\newblock


\bibitem[\protect\citeauthoryear{Schneider, Gerstenberger, and
  Hoefler}{Schneider et~al\mbox{.}}{2012}]%
        {mpi-ddt-benchmark}
\bibfield{author}{\bibinfo{person}{T. Schneider}, \bibinfo{person}{R.
  Gerstenberger}, {and} \bibinfo{person}{T. Hoefler}.}
  \bibinfo{year}{2012}\natexlab{}.
\newblock \showarticletitle{{Micro-Applications for Communication Data Access
  Patterns and MPI Datatypes}}. In \bibinfo{booktitle}{{\em Recent Advances in
  the Message Passing Interface - Proceedings of the 19th European MPI Users'
  Group Meeting, EuroMPI 2012, 2012.}}, Vol.~\bibinfo{volume}{7490}.
  \bibinfo{publisher}{Springer}, \bibinfo{pages}{121--131}.
\newblock
\showISBNx{978-3-642-33517-4}


\bibitem[\protect\citeauthoryear{Schneider, Gerstenberger, and
  Hoefler}{Schneider et~al\mbox{.}}{2014}]%
        {schneider-app-oriented-ping-pong}
\bibfield{author}{\bibinfo{person}{T. Schneider}, \bibinfo{person}{R.
  Gerstenberger}, {and} \bibinfo{person}{T. Hoefler}.}
  \bibinfo{year}{2014}\natexlab{}.
\newblock \showarticletitle{{Application-oriented ping-pong benchmarking: how
  to assess the real communication overheads}}.
\newblock \bibinfo{journal}{{\em Journal of Computing\/}} \bibinfo{volume}{96},
  \bibinfo{number}{4} (\bibinfo{date}{Apr.} \bibinfo{year}{2014}),
  \bibinfo{pages}{279--292}.
\newblock
\showISSN{0010-485X}


\bibitem[\protect\citeauthoryear{Hoefler and Gottlieb}{Hoefler and
  Gottlieb}{2010}]%
        {hoefler-datatypes}
\bibfield{author}{\bibinfo{person}{T. Hoefler} {and} \bibinfo{person}{S.
  Gottlieb}.} \bibinfo{year}{2010}\natexlab{}.
\newblock \showarticletitle{{Parallel Zero-Copy Algorithms for Fast Fourier
  Transform and Conjugate Gradient using MPI Datatypes}}. In
  \bibinfo{booktitle}{{\em Recent Advances in the Message Passing Interface
  (EuroMPI'10)}}, Vol.~\bibinfo{volume}{LNCS 6305}.
  \bibinfo{publisher}{Springer}, \bibinfo{pages}{132--141}.
\newblock
\showISBNx{078-3-642-15645-8}
\showISSN{0302-9743}


\bibitem[\protect\citeauthoryear{Gropp, Hoefler, Thakur, and Traeff}{Gropp
  et~al\mbox{.}}{2011}]%
        {gropp-datatype-performance}
\bibfield{author}{\bibinfo{person}{W. Gropp}, {et~al\mbox{.}}}
  \bibinfo{year}{2011}\natexlab{}.
\newblock \showarticletitle{{Performance Expectations and Guidelines for MPI
  Derived Datatypes}}. In \bibinfo{booktitle}{{\em Recent Advances in the
  Message Passing Interface (EuroMPI'11)}}, Vol.~\bibinfo{volume}{6960}.
  \bibinfo{publisher}{Springer}, \bibinfo{pages}{150--159}.
\newblock
\showISBNx{978-3-642-24448-3}


\bibitem[\protect\citeauthoryear{Schneider, Kjolstad, and Hoefler}{Schneider
  et~al\mbox{.}}{2013}]%
        {schneider-rtcompmpiddt}
\bibfield{author}{\bibinfo{person}{T. Schneider}, \bibinfo{person}{F.
  Kjolstad}, {and} \bibinfo{person}{T. Hoefler}.}
  \bibinfo{year}{2013}\natexlab{}.
\newblock \showarticletitle{{MPI Datatype Processing using Runtime
  Compilation}}. In \bibinfo{booktitle}{{\em Proceedings of the 20th European
  MPI Users' Group Meeting}}. \bibinfo{publisher}{ACM},
  \bibinfo{pages}{19--24}.
\newblock
\showISBNx{978-1-4503-1903-4}


\bibitem[\protect\citeauthoryear{Santhanaraman, Wu, and Panda}{Santhanaraman
  et~al\mbox{.}}{2004}]%
        {panda1}
\bibfield{author}{\bibinfo{person}{G. Santhanaraman}, \bibinfo{person}{J. Wu},
  {and} \bibinfo{person}{D.~K Panda}.} \bibinfo{year}{2004}\natexlab{}.
\newblock \showarticletitle{Zero-copy MPI derived datatype communication over
  InfiniBand}. In \bibinfo{booktitle}{{\em European Parallel Virtual
  Machine/Message Passing Interface Users' Group Meeting}}. Springer,
  \bibinfo{pages}{47--56}.
\newblock


\bibitem[\protect\citeauthoryear{Wang, Potluri, Luo, Singh, Ouyang, Sur, and
  Panda}{Wang et~al\mbox{.}}{2011}]%
        {panda2}
\bibfield{author}{\bibinfo{person}{H. Wang}, {et~al\mbox{.}}}
  \bibinfo{year}{2011}\natexlab{}.
\newblock \showarticletitle{Optimized non-contiguous MPI datatype communication
  for GPU clusters: Design, implementation and evaluation with MVAPICH2}. In
  \bibinfo{booktitle}{{\em 2011 IEEE International Conference on Cluster
  Computing}}. IEEE, \bibinfo{pages}{308--316}.
\newblock


\bibitem[\protect\citeauthoryear{Hoefler, Girolamo, Taranov, Grant, and
  Brightwell}{Hoefler et~al\mbox{.}}{2017}]%
        {spin}
\bibfield{author}{\bibinfo{person}{T. Hoefler}, {et~al\mbox{.}}}
  \bibinfo{year}{2017}\natexlab{}.
\newblock \showarticletitle{{sPIN: High-performance streaming Processing in the
  Network}}. In \bibinfo{booktitle}{{\em Proceedings of the International
  Conference for High Performance Computing, Networking, Storage and Analysis
  (SC17)}}.
\newblock


\bibitem[\protect\citeauthoryear{Van~der Wijngaart and Wong}{Van~der Wijngaart
  and Wong}{2002}]%
        {van2002parallel}
\bibfield{author}{\bibinfo{person}{R.~F Van~der Wijngaart} {and}
  \bibinfo{person}{P. Wong}.} \bibinfo{year}{2002}\natexlab{}.
\newblock \showarticletitle{NAS parallel benchmarks version 2.4}.
\newblock  (\bibinfo{year}{2002}).
\newblock


\bibitem[\protect\citeauthoryear{Nieplocha, Tipparaju, Krishnan, and
  Panda}{Nieplocha et~al\mbox{.}}{2006}]%
        {nieplocha2006high}
\bibfield{author}{\bibinfo{person}{J. Nieplocha}, {et~al\mbox{.}}}
  \bibinfo{year}{2006}\natexlab{}.
\newblock \showarticletitle{High performance remote memory access
  communication: The ARMCI approach}.
\newblock \bibinfo{journal}{{\em The International Journal of High Performance
  Computing Applications\/}} \bibinfo{volume}{20}, \bibinfo{number}{2}
  (\bibinfo{year}{2006}), \bibinfo{pages}{233--253}.
\newblock


\bibitem[\protect\citeauthoryear{Chapman, Curtis, Pophale, Poole, Kuehn,
  Koelbel, and Smith}{Chapman et~al\mbox{.}}{2010}]%
        {chapman2010introducing}
\bibfield{author}{\bibinfo{person}{B. Chapman}, {et~al\mbox{.}}}
  \bibinfo{year}{2010}\natexlab{}.
\newblock \showarticletitle{Introducing OpenSHMEM: SHMEM for the PGAS
  community}. In \bibinfo{booktitle}{{\em Proceedings of the Fourth Conference
  on Partitioned Global Address Space Programming Model}}. ACM,
  \bibinfo{pages}{2}.
\newblock


\bibitem[\protect\citeauthoryear{Mellor-Crummey, Adhianto, Scherer~III, and
  Jin}{Mellor-Crummey et~al\mbox{.}}{2009}]%
        {mellor2009new}
\bibfield{author}{\bibinfo{person}{J. Mellor-Crummey}, {et~al\mbox{.}}}
  \bibinfo{year}{2009}\natexlab{}.
\newblock \showarticletitle{A new vision for Coarray Fortran}. In
  \bibinfo{booktitle}{{\em Proceedings of the Third Conference on Partitioned
  Global Address Space Programing Models}}. ACM, \bibinfo{pages}{5}.
\newblock


\bibitem[\protect\citeauthoryear{El-Ghazawi and Smith}{El-Ghazawi and
  Smith}{2006}]%
        {el2006upc}
\bibfield{author}{\bibinfo{person}{T. El-Ghazawi} {and} \bibinfo{person}{L.
  Smith}.} \bibinfo{year}{2006}\natexlab{}.
\newblock \showarticletitle{UPC: unified parallel C}. In
  \bibinfo{booktitle}{{\em Proceedings of the 2006 ACM/IEEE conference on
  Supercomputing}}. ACM, \bibinfo{pages}{27}.
\newblock


\bibitem[\protect\citeauthoryear{Forum}{Forum}{2012}]%
        {mpi-3.0}
\bibfield{author}{\bibinfo{person}{Message Passing~Interface Forum}.}
  \bibinfo{year}{2012}\natexlab{}.
\newblock \bibinfo{title}{{MPI: A Message-Passing Interface Standard Version
  3.0}}.
\newblock   (\bibinfo{date}{09} \bibinfo{year}{2012}).
\newblock
\newblock
\shownote{Chapter author for Collective Communication, Process Topologies, and
  One Sided Communications.}


\bibitem[\protect\citeauthoryear{Gropp, Lusk, and Swider}{Gropp
  et~al\mbox{.}}{1999}]%
        {gropp1999improving}
\bibfield{author}{\bibinfo{person}{W. Gropp}, \bibinfo{person}{E. Lusk}, {and}
  \bibinfo{person}{D. Swider}.} \bibinfo{year}{1999}\natexlab{}.
\newblock \showarticletitle{Improving the performance of MPI derived
  datatypes}. In \bibinfo{booktitle}{{\em Proceedings of the Third MPI
  Developer's and User's Conference}}. MPI Software Technology Press,
  \bibinfo{pages}{25--30}.
\newblock


\bibitem[\protect\citeauthoryear{Byna, Sun, Thakur, and Gropp}{Byna
  et~al\mbox{.}}{2006}]%
        {byna2006automatic}
\bibfield{author}{\bibinfo{person}{S. Byna}, {et~al\mbox{.}}}
  \bibinfo{year}{2006}\natexlab{}.
\newblock \showarticletitle{Automatic memory optimizations for improving MPI
  derived datatype performance}. In \bibinfo{booktitle}{{\em European Parallel
  Virtual Machine/Message Passing Interface Users' Group Meeting}}. Springer,
  \bibinfo{pages}{238--246}.
\newblock


\bibitem[\protect\citeauthoryear{Tanabe and Nakajo}{Tanabe and Nakajo}{2008}]%
        {tanabe2008introduction}
\bibfield{author}{\bibinfo{person}{N. Tanabe} {and} \bibinfo{person}{H.
  Nakajo}.} \bibinfo{year}{2008}\natexlab{}.
\newblock \showarticletitle{Introduction to acceleration for MPI derived
  datatypes using an enhancer of memory and network}. In
  \bibinfo{booktitle}{{\em European Parallel Virtual Machine/Message Passing
  Interface Users' Group Meeting}}. Springer, \bibinfo{pages}{324--325}.
\newblock


\bibitem[\protect\citeauthoryear{Tr\"{a}ff}{Tr\"{a}ff}{2014}]%
        {Traff:2014:OMD:2642769.2642771}
\bibfield{author}{\bibinfo{person}{J.~L. Tr\"{a}ff}.}
  \bibinfo{year}{2014}\natexlab{}.
\newblock \showarticletitle{Optimal MPI Datatype Normalization for Vector and
  Index-block Types}. In \bibinfo{booktitle}{{\em Proceedings of the 21st
  European MPI Users' Group Meeting}} {\em (\bibinfo{series}{EuroMPI/ASIA
  '14})}. \bibinfo{publisher}{ACM}, \bibinfo{address}{New York, NY, USA},
  Article \bibinfo{articleno}{33}, \bibinfo{numpages}{6}~pages.
\newblock
\showISBNx{978-1-4503-2875-3}
\showDOI{%
\url{http://dx.doi.org/10.1145/2642769.2642771}}


\bibitem[\protect\citeauthoryear{Ross, Latham, Gropp, Lusk, and Thakur}{Ross
  et~al\mbox{.}}{2009}]%
        {ross2009processing}
\bibfield{author}{\bibinfo{person}{R. Ross}, {et~al\mbox{.}}}
  \bibinfo{year}{2009}\natexlab{}.
\newblock \showarticletitle{Processing MPI datatypes outside MPI}. In
  \bibinfo{booktitle}{{\em European Parallel Virtual Machine/Message Passing
  Interface Users' Group Meeting}}. Springer, \bibinfo{pages}{42--53}.
\newblock


\bibitem[\protect\citeauthoryear{Ross, Miller, and Gropp}{Ross
  et~al\mbox{.}}{2003}]%
        {ross2003implementing}
\bibfield{author}{\bibinfo{person}{R. Ross}, \bibinfo{person}{N. Miller}, {and}
  \bibinfo{person}{W.~D Gropp}.} \bibinfo{year}{2003}\natexlab{}.
\newblock \showarticletitle{Implementing fast and reusable datatype
  processing}. In \bibinfo{booktitle}{{\em European Parallel Virtual
  Machine/Message Passing Interface Users' Group Meeting}}. Springer,
  \bibinfo{pages}{404--413}.
\newblock


\bibitem[\protect\citeauthoryear{Kurth, Vogel, Capotondi, Marongiu, and
  Benini}{Kurth et~al\mbox{.}}{2017}]%
        {kurth2017hero}
\bibfield{author}{\bibinfo{person}{A. Kurth}, {et~al\mbox{.}}}
  \bibinfo{year}{2017}\natexlab{}.
\newblock \showarticletitle{HERO: Heterogeneous embedded research platform for
  exploring RISC-V manycore accelerators on FPGA}.
\newblock \bibinfo{journal}{{\em arXiv preprint arXiv:1712.06497\/}}
  (\bibinfo{year}{2017}).
\newblock


\bibitem[\protect\citeauthoryear{{Rossi}, {Pullini}, {Loi}, {Gautschi},
  {Gürkaynak}, {Teman}, {Constantin}, {Burg}, {Miro-Panades}, {Beignè},
  {Clermidy}, {Flatresse}, and {Benini}}{{Rossi} et~al\mbox{.}}{2017}]%
        {Rossi2017}
\bibfield{author}{\bibinfo{person}{D. {Rossi}}, {et~al\mbox{.}}}
  \bibinfo{year}{2017}\natexlab{}.
\newblock \showarticletitle{Energy-Efficient Near-Threshold Parallel Computing:
  The PULPv2 Cluster}.
\newblock \bibinfo{journal}{{\em IEEE Micro\/}} \bibinfo{volume}{37},
  \bibinfo{number}{5} (\bibinfo{date}{Sep.} \bibinfo{year}{2017}),
  \bibinfo{pages}{20--31}.
\newblock


\bibitem[\protect\citeauthoryear{{Gautschi}, {Schiavone}, {Traber}, {Loi},
  {Pullini}, {Rossi}, {Flamand}, {Gürkaynak}, and {Benini}}{{Gautschi}
  et~al\mbox{.}}{2017}]%
        {gautschi2017ri5cy}
\bibfield{author}{\bibinfo{person}{M. {Gautschi}}, {et~al\mbox{.}}}
  \bibinfo{year}{2017}\natexlab{}.
\newblock \showarticletitle{Near-Threshold RISC-V Core With DSP Extensions for
  Scalable IoT Endpoint Devices}.
\newblock \bibinfo{journal}{{\em IEEE Transactions on Very Large Scale
  Integration (VLSI) Systems\/}} \bibinfo{volume}{25}, \bibinfo{number}{10}
  (\bibinfo{date}{Oct} \bibinfo{year}{2017}), \bibinfo{pages}{2700--2713}.
\newblock
\showISSN{1063-8210}
\showDOI{%
\url{http://dx.doi.org/10.1109/TVLSI.2017.2654506}}


\bibitem[\protect\citeauthoryear{Technologies}{Technologies}{2019}]%
        {blueField2019}
\bibfield{author}{\bibinfo{person}{Mellanox Technologies}.}
  \bibinfo{year}{2019}\natexlab{}.
\newblock \bibinfo{title}{Mellanox BlueField SmartNIC}.
\newblock
  \bibinfo{howpublished}{\url{http://www.mellanox.com/related-docs/prod_adapter_cards/PB_BlueField_Smart_NIC.pdf}}.
    (\bibinfo{year}{2019}).
\newblock
\newblock
\shownote{Online; accessed 05. April 2019.}


\bibitem[\protect\citeauthoryear{{Mair}, {Gammie}, {Wang}, {Lagerquist},
  {Chung}, {Gururajarao}, {Kao}, {Rajagopalan}, {Saha}, {Jain}, {Wang},
  {Ouyang}, {Wen}, {Thippana}, {Chen}, {Rahman}, {Chau}, {Varma}, {Flachs},
  {Peng}, {Tsai}, {Lin}, {Fu}, {Kuo}, {Yong}, {Peng}, {Shieh}, {Wu}, and
  {Ko}}{{Mair} et~al\mbox{.}}{2016}]%
        {Mair2016}
\bibfield{author}{\bibinfo{person}{H.~T. {Mair}}, {et~al\mbox{.}}}
  \bibinfo{year}{2016}\natexlab{}.
\newblock \showarticletitle{4.3 A 20nm 2.5GHz ultra-low-power tri-cluster CPU
  subsystem with adaptive power allocation for optimal mobile SoC performance}.
  In \bibinfo{booktitle}{{\em IEEE International Solid-State Circuits
  Conference (ISSCC)}}. \bibinfo{pages}{76--77}.
\newblock


\bibitem[\protect\citeauthoryear{{Pyo}, {Shin}, {Lee}, {Bae}, {Kim}, {Kim},
  {Shin}, {Kwon}, {Oh}, {Lim}, {Lee}, {Lee}, {Hong}, {Chae}, {Lee}, {Lee},
  {Song}, {Kim}, {Park}, {Kim}, {Yun}, {Cho}, {Son}, and {Park}}{{Pyo}
  et~al\mbox{.}}{2015}]%
        {Pyo2015}
\bibfield{author}{\bibinfo{person}{J. {Pyo}}, {et~al\mbox{.}}}
  \bibinfo{year}{2015}\natexlab{}.
\newblock \showarticletitle{23.1 20nm high-K metal-gate heterogeneous 64b
  quad-core CPUs and hexa-core GPU for high-performance and energy-efficient
  mobile application processor}. In \bibinfo{booktitle}{{\em 2015 IEEE
  International Solid-State Circuits Conference - (ISSCC) Digest of Technical
  Papers}}. \bibinfo{pages}{1--3}.
\newblock


\bibitem[\protect\citeauthoryear{{Sohan}, {Rice}, {Andrew}, and
  {Mansley}}{{Sohan} et~al\mbox{.}}{2010}]%
        {Sohan2010}
\bibfield{author}{\bibinfo{person}{R. {Sohan}}, {et~al\mbox{.}}}
  \bibinfo{year}{2010}\natexlab{}.
\newblock \showarticletitle{Characterizing 10 Gbps network interface energy
  consumption}. In \bibinfo{booktitle}{{\em IEEE Local Computer Network
  Conference}}. \bibinfo{pages}{268--271}.
\newblock


\bibitem[\protect\citeauthoryear{Janssen, Adalsteinsson, Cranford, Kenny,
  Pinar, Evensky, and Mayo}{Janssen et~al\mbox{.}}{2010}]%
        {janssen2010simulator}
\bibfield{author}{\bibinfo{person}{C.~L Janssen}, {et~al\mbox{.}}}
  \bibinfo{year}{2010}\natexlab{}.
\newblock \showarticletitle{A simulator for large-scale parallel computer
  architectures}.
\newblock \bibinfo{journal}{{\em International Journal of Distributed Systems
  and Technologies (IJDST)\/}} \bibinfo{volume}{1}, \bibinfo{number}{2}
  (\bibinfo{year}{2010}), \bibinfo{pages}{57--73}.
\newblock


\bibitem[\protect\citeauthoryear{Binkert, Beckmann, Black, Reinhardt, Saidi,
  Basu, Hestness, Hower, Krishna, Sardashti, et~al\mbox{.}}{Binkert
  et~al\mbox{.}}{2011}]%
        {binkert2011gem5}
\bibfield{author}{\bibinfo{person}{Nathan Binkert}, {et~al\mbox{.}}}
  \bibinfo{year}{2011}\natexlab{}.
\newblock \showarticletitle{The gem5 simulator}.
\newblock \bibinfo{journal}{{\em ACM SIGARCH Computer Architecture News\/}}
  \bibinfo{volume}{39}, \bibinfo{number}{2} (\bibinfo{year}{2011}),
  \bibinfo{pages}{1--7}.
\newblock


\bibitem[\protect\citeauthoryear{Endo, Courouss{\'e}, and Charles}{Endo
  et~al\mbox{.}}{2014}]%
        {endo2014micro}
\bibfield{author}{\bibinfo{person}{F.~A Endo}, \bibinfo{person}{D.
  Courouss{\'e}}, {and} \bibinfo{person}{H. Charles}.}
  \bibinfo{year}{2014}\natexlab{}.
\newblock \showarticletitle{Micro-architectural simulation of in-order and
  out-of-order arm microprocessors with gem5}. In \bibinfo{booktitle}{{\em 2014
  International Conference on Embedded Computer Systems: Architectures,
  Modeling, and Simulation (SAMOS XIV)}}. IEEE, \bibinfo{pages}{266--273}.
\newblock


\bibitem[\protect\citeauthoryear{Tousi and Zhu}{Tousi and Zhu}{2017}]%
        {tousi2017arm}
\bibfield{author}{\bibinfo{person}{A. Tousi} {and} \bibinfo{person}{C. Zhu}.}
  \bibinfo{year}{2017}\natexlab{}.
\newblock \bibinfo{title}{Arm Research Starter Kit: System Modeling using
  gem5}.
\newblock   (\bibinfo{year}{2017}).
\newblock


\bibitem[\protect\citeauthoryear{Hoefler, Schneider, and Lumsdaine}{Hoefler
  et~al\mbox{.}}{2010}]%
        {hoefler-loggopsim}
\bibfield{author}{\bibinfo{person}{T. Hoefler}, \bibinfo{person}{T. Schneider},
  {and} \bibinfo{person}{A. Lumsdaine}.} \bibinfo{year}{2010}\natexlab{}.
\newblock \showarticletitle{{LogGOPSim - Simulating Large-Scale Applications in
  the LogGOPS Model}}. In \bibinfo{booktitle}{{\em Proceedings of the 19th ACM
  International Symposium on High Performance Distributed Computing}}.
  \bibinfo{publisher}{ACM}, \bibinfo{pages}{597--604}.
\newblock
\showISBNx{978-1-60558-942-8}


\bibitem[\protect\citeauthoryear{Laboratory}{Laboratory}{2018}]%
        {comb}
\bibfield{author}{\bibinfo{person}{Lawrence Livermore~National Laboratory}.}
  \bibinfo{year}{2018}\natexlab{}.
\newblock \showarticletitle{Comb is a communication performance benchmarking
  tool.}
\newblock  (\bibinfo{year}{2018}).
\newblock
\showURL{%
\url{https://github.com/LLNL/Comb}}


\bibitem[\protect\citeauthoryear{Plimpton}{Plimpton}{1995}]%
        {Plimpton:1995:FPA:201627.201628}
\bibfield{author}{\bibinfo{person}{S. Plimpton}.}
  \bibinfo{year}{1995}\natexlab{}.
\newblock \showarticletitle{Fast Parallel Algorithms for Short-range Molecular
  Dynamics}.
\newblock \bibinfo{journal}{{\em J. Comput. Phys.\/}} \bibinfo{volume}{117},
  \bibinfo{number}{1} (\bibinfo{date}{March} \bibinfo{year}{1995}),
  \bibinfo{pages}{1--19}.
\newblock
\showISSN{0021-9991}
\showDOI{%
\url{http://dx.doi.org/10.1006/jcph.1995.1039}}


\bibitem[\protect\citeauthoryear{Bernard, Ogilvie, DeGrand, DeTar, Gottlieb,
  Krasnitz, Sugar, and Toussaint}{Bernard et~al\mbox{.}}{1991}]%
        {bernard1991studying}
\bibfield{author}{\bibinfo{person}{C. Bernard}, {et~al\mbox{.}}}
  \bibinfo{year}{1991}\natexlab{}.
\newblock \showarticletitle{Studying quarks and gluons on MIMD parallel
  computers}.
\newblock \bibinfo{journal}{{\em The International Journal of Supercomputing
  Applications\/}} \bibinfo{volume}{5}, \bibinfo{number}{4}
  (\bibinfo{year}{1991}), \bibinfo{pages}{61--70}.
\newblock


\bibitem[\protect\citeauthoryear{Carrington, Komatitsch, Laurenzano, Tikir,
  Mich{\'e}a, Le~Goff, Snavely, and Tromp}{Carrington et~al\mbox{.}}{2008}]%
        {Carrington:2008:HSG:1413370.1413432}
\bibfield{author}{\bibinfo{person}{L. Carrington}, {et~al\mbox{.}}}
  \bibinfo{year}{2008}\natexlab{}.
\newblock \showarticletitle{High-frequency Simulations of Global Seismic Wave
  Propagation Using SPECFEM3D GLOBE on 62K Processors}. In
  \bibinfo{booktitle}{{\em Proceedings of the 2008 ACM/IEEE Conference on
  Supercomputing}} {\em (\bibinfo{series}{SC '08})}. \bibinfo{publisher}{IEEE
  Press}, \bibinfo{address}{Piscataway, NJ, USA}, Article
  \bibinfo{articleno}{60}, \bibinfo{numpages}{11}~pages.
\newblock
\showISBNx{978-1-4244-2835-9}
\showURL{%
\url{http://dl.acm.org/citation.cfm?id=1413370.1413432}}


\bibitem[\protect\citeauthoryear{Sjogreen}{Sjogreen}{2018}]%
        {sw4lite}
\bibfield{author}{\bibinfo{person}{B Sjogreen}.}
  \bibinfo{year}{2018}\natexlab{}.
\newblock \bibinfo{booktitle}{{\em SW4 final report for iCOE}}.
\newblock \bibinfo{type}{{T}echnical {R}eport}. \bibinfo{institution}{Lawrence
  Livermore National Lab.(LLNL), Livermore, CA (United States)}.
\newblock


\bibitem[\protect\citeauthoryear{Skamarock and Klemp}{Skamarock and
  Klemp}{2008}]%
        {Skamarock:2008:TNA:1347465.1347775}
\bibfield{author}{\bibinfo{person}{W.~C. Skamarock} {and}
  \bibinfo{person}{J.~B. Klemp}.} \bibinfo{year}{2008}\natexlab{}.
\newblock \showarticletitle{A Time-split Nonhydrostatic Atmospheric Model for
  Weather Research and Forecasting Applications}.
\newblock \bibinfo{journal}{{\em J. Comput. Phys.\/}} \bibinfo{volume}{227},
  \bibinfo{number}{7} (\bibinfo{date}{March} \bibinfo{year}{2008}),
  \bibinfo{pages}{3465--3485}.
\newblock
\showISSN{0021-9991}
\showDOI{%
\url{http://dx.doi.org/10.1016/j.jcp.2007.01.037}}


\bibitem[\protect\citeauthoryear{Neugebauer, Antichi, Zazo, Audzevich,
  L{\'o}pez-Buedo, and Moore}{Neugebauer et~al\mbox{.}}{2018}]%
        {neugebauer2018understanding}
\bibfield{author}{\bibinfo{person}{R. Neugebauer}, {et~al\mbox{.}}}
  \bibinfo{year}{2018}\natexlab{}.
\newblock \showarticletitle{Understanding PCIe performance for end host
  networking}. In \bibinfo{booktitle}{{\em Proceedings of the 2018 Conference
  of the ACM Special Interest Group on Data Communication}}. ACM,
  \bibinfo{pages}{327--341}.
\newblock


\bibitem[\protect\citeauthoryear{Martinasso, Kwasniewski, Alam, Shulthess, and
  Hoefler}{Martinasso et~al\mbox{.}}{2016}]%
        {pciecong}
\bibfield{author}{\bibinfo{person}{M. Martinasso}, {et~al\mbox{.}}}
  \bibinfo{year}{2016}\natexlab{}.
\newblock \showarticletitle{{A PCIe Congestion-Aware Performance Model for
  Densely Populated Accelerator Servers}}. In \bibinfo{booktitle}{{\em
  Proceedings of the International Conference for High Performance Computing,
  Networking, Storage and Analysis (SC16)}}. \bibinfo{publisher}{IEEE Press},
  \bibinfo{pages}{63:1--63:11}.
\newblock
\showISBNx{978-1-4673-8815-3}


\bibitem[\protect\citeauthoryear{Hoefler, Siebert, and Lumsdaine}{Hoefler
  et~al\mbox{.}}{2009}]%
        {hoefler-goal}
\bibfield{author}{\bibinfo{person}{T. Hoefler}, \bibinfo{person}{C. Siebert},
  {and} \bibinfo{person}{A. Lumsdaine}.} \bibinfo{year}{2009}\natexlab{}.
\newblock \showarticletitle{{Group Operation Assembly Language - A Flexible Way
  to Express Collective Communication}}, In \bibinfo{booktitle}{ICPP-2009 - The
  38th International Conference on Parallel Processing}.  (\bibinfo{date}{Sep.}
  \bibinfo{year}{2009}).
\newblock
\showISBNx{978-0-7695-3802-0}


\bibitem[\protect\citeauthoryear{Tr{\"a}ff, Hempel, Ritzdorf, and
  Zimmermann}{Tr{\"a}ff et~al\mbox{.}}{1999}]%
        {10.1007/3-540-48158-3_14}
\bibfield{author}{\bibinfo{person}{J.~L. Tr{\"a}ff}, {et~al\mbox{.}}}
  \bibinfo{year}{1999}\natexlab{}.
\newblock \showarticletitle{Flattening on the Fly: efficient handling of MPI
  derived datatypes}. In \bibinfo{booktitle}{{\em Recent Advances in Parallel
  Virtual Machine and Message Passing Interface}},
  \bibfield{editor}{\bibinfo{person}{Jack Dongarra}, \bibinfo{person}{Emilio
  Luque}, {and} \bibinfo{person}{Tom{\`a}s Margalef}} (Eds.).
  \bibinfo{publisher}{Springer Berlin Heidelberg}, \bibinfo{address}{Berlin,
  Heidelberg}, \bibinfo{pages}{109--116}.
\newblock
\showISBNx{978-3-540-48158-4}


\bibitem[\protect\citeauthoryear{Prabhu and Gropp}{Prabhu and Gropp}{2015}]%
        {prabhu2015dame}
\bibfield{author}{\bibinfo{person}{T. Prabhu} {and} \bibinfo{person}{W.
  Gropp}.} \bibinfo{year}{2015}\natexlab{}.
\newblock \showarticletitle{DAME: A runtime-compiled engine for derived
  datatypes}. In \bibinfo{booktitle}{{\em Proceedings of the 22nd European MPI
  Users' Group Meeting}}. ACM, \bibinfo{pages}{4}.
\newblock


\bibitem[\protect\citeauthoryear{Schneider, Gerstenberger, and
  Hoefler}{Schneider et~al\mbox{.}}{2013}]%
        {schneider2013compiler}
\bibfield{author}{\bibinfo{person}{T. Schneider}, \bibinfo{person}{R.
  Gerstenberger}, {and} \bibinfo{person}{T. Hoefler}.}
  \bibinfo{year}{2013}\natexlab{}.
\newblock \showarticletitle{Compiler optimizations for non-contiguous remote
  data movement}. In \bibinfo{booktitle}{{\em International Workshop on
  Languages and Compilers for Parallel Computing}}. Springer,
  \bibinfo{pages}{307--321}.
\newblock


\bibitem[\protect\citeauthoryear{Eicken, Culler, Goldstein, and
  Schauser}{Eicken et~al\mbox{.}}{1992}]%
        {eicken1992active}
\bibfield{author}{\bibinfo{person}{TV Eicken}, {et~al\mbox{.}}}
  \bibinfo{year}{1992}\natexlab{}.
\newblock \showarticletitle{Active messages: a mechanism for integrated
  communication and computation}. In \bibinfo{booktitle}{{\em [1992]
  Proceedings the 19th Annual International Symposium on Computer
  Architecture}}. IEEE, \bibinfo{pages}{256--266}.
\newblock


\bibitem[\protect\citeauthoryear{Besta and Hoefler}{Besta and Hoefler}{2015a}]%
        {besta-aam}
\bibfield{author}{\bibinfo{person}{M. Besta} {and} \bibinfo{person}{T.
  Hoefler}.} \bibinfo{year}{2015}\natexlab{a}.
\newblock \showarticletitle{{Accelerating Irregular Computations with Hardware
  Transactional Memory and Active Messages}}. In \bibinfo{booktitle}{{\em
  Proceedings of the 24th Symposium on High-Performance Parallel and
  Distributed Computing (HPDC'15)}}. \bibinfo{publisher}{ACM},
  \bibinfo{pages}{161--172}.
\newblock
\showISBNx{978-1-4503-3550-8}


\bibitem[\protect\citeauthoryear{Besta and Hoefler}{Besta and Hoefler}{2015b}]%
        {besta-aa}
\bibfield{author}{\bibinfo{person}{M. Besta} {and} \bibinfo{person}{T.
  Hoefler}.} \bibinfo{year}{2015}\natexlab{b}.
\newblock \showarticletitle{{Active Access: A Mechanism for High-Performance
  Distributed Data-Centric Computations}}. In \bibinfo{booktitle}{{\em
  Proceedings of the 29th International Conference on Supercomputing
  (ICS'15)}}. \bibinfo{publisher}{ACM}, \bibinfo{pages}{155--164}.
\newblock
\showISBNx{978-1-4503-3559-1}


\end{thebibliography}

\clearpage









\end{document}